\documentclass[notitlepage,a4paper,aps,prd,onecolumn,superscriptaddress,nofootinbib,groupedaddress]{revtex4}

\usepackage{amsmath}
\usepackage{amsfonts,color}
\usepackage{amsthm}
\usepackage{pdfpages}
\usepackage{empheq}
\usepackage{amssymb,float}
\usepackage{booktabs,multirow}
\usepackage{titlesec}
\usepackage{hyphenat}
\usepackage{ragged2e}
\usepackage{amsmath}
\usepackage{accents}
\usepackage[utf8]{inputenc}
\usepackage{color}
\usepackage{hyperref}
\usepackage{enumitem}
\usepackage{tikz}
\usetikzlibrary{shapes.geometric}
\usetikzlibrary{arrows.meta,arrows}

\titleformat{\paragraph}
{\normalfont\normalsize\bfseries}{\theparagraph}{1em}{}
\titlespacing*{\paragraph}
{0pt}{3.25ex plus 1ex minus .2ex}{1.5ex plus .2ex}
\newcommand{\lc}[1]{\accentset{\circ}{#1}}
\allowdisplaybreaks
\hypersetup{
    colorlinks=true,
    linkcolor=blue,
    filecolor=magenta,      
    citecolor=blue
}

\begin{document}

\title{Dirac-Bergmann analysis and Degrees of Freedom of Coincident $f(Q)$-gravity}
\author{Kyosuke {\sc Tomonari}}
\email{tomonari.k.aa@m.titech.ac.jp}
\email{ktomonari.phys@gmail.com}
\affiliation{Department of Physics, Tokyo Institute of Technology, 2-12-1 Ookayama, Meguro-ku, Tokyo 152-8551, Japan}
\author{Sebastian Bahamonde}
\email{sbahamondebeltran@gmail.com}
\affiliation{Kavli Institute for the Physics and Mathematics of the Universe (Kavli IPMU), The University of Tolyo, 5-1-5 Kashiwanoha, Kashiwa, Chiba, 277-8583, Japan}

\begin{abstract}
We investigate the propagating degrees of freedom of $f(Q)$-gravity in a $4$-dimensional space-time under the imposition of the coincident gauge by performing the Dirac-Bergmann analysis. In this work, we start with a top-down reconstruction of the metric-affine gauge theory of gravity based only on the concept of a vector bundle. Then, the so-called geometrical trinity of gravity is introduced and the role of the coincident GR is clarified. After that, we reveal relationships between the boundary terms in the variational principle and the symplectic structure of the theory in order to confirm the validity of the analysis for our studied theories. Then, as examples, we revisit the analysis of GR and its $f(\lc{R})$-extensions. Finally, after reviewing the Dirac-Bergmann analysis of the coincident GR and that of $f(T)$-gravity, we perform the analysis of coincident $f(Q)$-gravity. Under the imposition of appropriate spatial boundary conditions, we find that, as a generic case, the theory has five primary, three secondary, and two tertiary constraint densities and all these constraint densities are classified into second-class constraint density; the number six is the propagating degrees of freedom of the theory and there are no longer any remaining gauge degrees of freedom. We also discuss the condition of providing seven pDoF as a generic case. The violation of diffeomorphism invariance of coincident $f(Q)$-gravity make it possible to emerge such several sectors.
\end{abstract}

\maketitle

\section{\label{sec:01}Introduction}
General Relativity (GR) is the most successful theory to describe the wide range of gravitational phenomena in terms of pseudo-Riemannian geometry based on Einstein's equivalence principle and the general covariance. However, from the physical point of view, there is no reason to restrict our theories to this particular geometry. In fact, Einstein himself reconstructed GR in an alternative way using another geometry based purely on torsion instead of curvature, labeled as Teleparallel gravity~\cite{Einstein1928}. For a detailed review on teleparallel gravity, see~\cite{Bahamonde:2021gfp}. In modern perspectives, it is known that GR has its equivalent formulation of the so-called geometrical trinity of gravity, in which gravitation is treated with the torsion (Teleparallel Equivalent to GR: TEGR) and/or the non-metricity (Symmetric Teleparallel Equivalent to GR: STEGR) instead of the curvature up to boundary terms~\cite{Nester:1998mp,Jimenez2019,Heisenberg:2018vsk}. Those two formalism assumes that the general curvature is vanishing. Furthermore, a non-linear extension of these equivalent formulations gives the emergences of their specific properties such as new propagating Degrees of Freedom (pDoF), breaking diffeomorphism and/or local Lorentz symmetry ({\it i\,.e\,.,} change of gauge Degrees of Freedom: gDoF), and departures in these formulations themselves~\cite{Buchdahl1970,Bahamonde:2021gfp,Jimenez2020}. In particular, the so-called $\lc{R}^{2}$-inflation model in cosmology, which is a class of $f(\lc{R})$-gravity, is one of the most suitable theories to explain inflation in the current observations~\cite{Planck18,Bessa2022}. That model is a good candidate for a consistent effective quantum gravity theory from the viewpoint of renormalization~\cite{Stelle1978,Starobinsky1980}. In order to investigate the pDoF of possible extensions of GR, TEGR, or STEGR, the Dirac-Bergmann analysis can be used~\cite{Dirac1950,Dirac1958,Bergmann1949,BergmannBrunings1949,Bergmann1950,AndersonBergmann1951}. 

In terms of the torsional sector (as in TEGR or their extensions), one can always formulate those theories in the so-called Weitzenb\"{o}ck gauge \cite{Weitzenboh1923} where the spin connection vanishes. In this context, the Dirac-Bergmann analysis of TEGR had already been completed and the structure as a constraint system had also been revealed out~\cite{Blagojevic2000,Maluf2001,Ferraro2016}. As expected, in a $4$-dimensional spacetime, TEGR has two pDoF, which is the same pDoF as GR, and the gauge symmetric structure, in other words, the Poisson Brackets algebra (PB-algebra) is also similar that the one of GR: the gDoF for the diffeomorphism symmetry is four~\cite{Blagojevic2000-2,Ferraro2016}. For the non-linear extension of TEGR, {\it i\,.e\,.,} the so-called ``{\it $f(T)$-gravity}'', the Dirac-Bergmann analysis had been performed~\cite{Li2011,Ong2013,Ferraro2018,Blagojevic2020}. In that case, the situation is different than $f(\lc{R})$-gravity \cite{Liang2017}; Furthermore, there were some controversies on the computation of the pDoF for that theory (see for example~\cite{Li2011,Ferraro2018,Blagojevic2020}). The authors in~\cite{Li2011} and~\cite{Blagojevic2000}, on one hand, state that $f(T)$-gravity has $n$ extra pDoF in a $(n+1)$-dimensional spacetime. On the other hand, the authors in~\cite{Ferraro2018} concluded that the extra pDoF is one in any spacetime dimension. One can check the details in~\cite{Blixt2021}. Other important viewpoints of these theories are that cosmological perturbations around flat and non-flat Friedmann–Lemaître–Robertson–Walker (FLRW) spacetimes have suggested the fact that this theory is infinitely strongly coupled for this spacetime~\cite{Golovnev:2018wbh,Bahamonde:2022ohm}. This means that $f(T)$-gravity cannot be used as a theory for cosmology as in the standard way, {\it i\,.e\,.,} using linear perturbation theory, since the new degrees of freedom are infinitely strongly coupled to the background and then linear perturbation theory breaks down. To unveil such a perspective, we have to know the exact pDoF of the theory, and the Dirac-Bergmann analysis plays a crucial role in achieving this purpose.

Recently, STEGR, the other sector of the geometrical trinity constructed from nonmetricity, has attained attention. This theory was initially constructed in~\cite{Nester:1998mp}, and further studied in~\cite{Jimenez2018,Jimenez2022,Blixt2023} by introducing the notion of the ``{\it coincident gauge}'' as an extra gauge freedom that one can always choose such that the connection is vanishing. Since this theory is equivalent to GR, it also has the same number of pDoF. Further, one can consider a non-linear extension of STEGR, such as the so-called ``{\it $f(Q)$-gravity}''. The Dirac-Bergmann analysis for that theory in the coincident gauge was performed in~\cite{Katsuragawa2022,Hu:2023gui}, where the authors argued that the pDoF is eight in $4$-dimensional spacetime. On the other hand, in the paper~\cite{Fabio2023}, the authors claimed that the Dirac-Bergman analysis breaks down for $f(Q)$-gravity, meaning that one cannot use this method to count the pDoF. Furthermore, the authors showed that the possible pDoF is up to seven. In our study, we pursue a possibility of pDoF being consistent with this range by using the Dirac-Bergmann analysis. In our opinion, this is a debatable point due to the technical point that one can assume that the spatial boundary terms in both the action and the PB-algebras can always be neglected by imposing appropriate spatial boundary conditions if it is necessary. This means that the second term of Eq (3.20) in~\cite{Fabio2023} does not give rise to any problematic terms at least theories that we will treat in the current paper. In addition, in the coincident $f(Q)$-gravity, since the diffeomorphism invariance are at least partly broken, it would be possible to generically exist a different pDoF for each class of coordinate systems, or equivalently, each class of ADM-foliations. Here, remark that the coincident $f(Q)$-gravity contains derivative terms up to first-order in the metric and dynamically totally different from higher-order derivative theories based on Riemannian geometry. For details, see Refs~\cite{Jimenez2019,Heisenberg:2018vsk}.

The construction of this paper is as follows: In Sec.~\ref{sec:02}, we introduce the gauge theory of gravity together with basic mathematical concepts to construct the metric-affine gauge theory of gravity and then give a short review of the geometrical trinity of gravity. In the context of the presented formulation, we also explain the coincident GR theory. In order to apply the Dirac-Bergmann analysis to field theories, one needs careful manipulations of boundary terms. In Sec.~\ref{sec:03}, we unveil that Gibbons-York-Hawking type boundary terms well-known in GR~\cite{York1972,GibbonsHawking1977,York1986,HawkingHorowitz1996} can be neglected without any change in the symplectic structure of a given system when performing the analysis. We also provide a prescription to circumvent the problematic situation concerning the PDEs of Lagrange multipliers which is mentioned in a series of works~\cite{Sundermeyer:1982,Blagojevic2020,Fabio2023}. Then we revisit the analysis of GR and $f(\lc{R})$-gravity to demonstrate how to work the statements declared in this section. In Sec.~\ref{sec:04}, we review the Dirac-Bergmann analysis of STEGR in the coincident gauge. In Sec.~\ref{sec:05}, after giving a brief review on $f(T)$-gravity to show a possibility for emerging several sectors and considering the role of the prescription, the analysis of coincident $f(Q)$-gravity is performed. We get the ${\rm pDoF}=6$ and ${\rm gDoF}=0$ in the generic case. Finally, in Sec.~\ref{sec:06}, we summarize this work with a discussion on the condition of providing seven pDoF as a generic sector and give future perspectives. 

Throughout this paper, we use units with $\kappa=c^{4}/16\pi G_{N}:=1$. In the Dirac-Bergmann analysis, we denote ``$\approx$'' as the weak equality~\cite{Dirac1950,Dirac1958} and ``$:\approx$'' as the imposition in the meaning of the weak equality. For quantities computed from the Levi-Civita connection, we use an over circle on top whereas, for a general connection, tildes are introduced. Also, Greek indices denote spacetime indices whereas small Latin ones, the tangent space indices. Capital Latin letters are introduced to distinguish the spatial indices in the ADM-foliation~\cite{ADM1959,ADM1960}.

\section{\label{sec:02}Metric-affine gauge theory of gravity with teleparallelism and Coincident GR}
In this section, we introduce the gauge approach for gravity and introduce the basic mathematical ingredients for that. Then, we introduce the metric-affine gauge theory of gravity and then give a short review of the geometrical trinity of gravity. In the context of the presented formulation, we also explain the so-called coincident GR.

\subsection{\label{sec:02:01}Gauge theories of gravity}
First of all, we introduce the fundamental mathematical objects to formulate gauge theories of gravity. Frame field (or vielbein) is a bundle isomorphism
between the tangent bundle $(T\mathcal{M},\mathcal{M},\pi)$ of a $(n+1)$-dimensional space-time manifold $\mathcal{M}$ and an internal space $(\mathcal{M}\times \mathbb{R}^{n+1},\mathcal{M},\rho)$, where $\pi$ and $\rho$ are diffeomorphisms from $T\mathcal{M}$ to $\mathcal{M}$ and from $\mathcal{M}\times \mathbb {R}^{n+1}$ to $\mathcal{M}$, respectively \cite{Baez1994,Nakahara2003}. That is, for an open set $U\subset \mathcal{M}$, ${\bf e}:\mathcal{M}\times \mathbb {R}^{n+1}\rightarrow T\mathcal{M}$ maps a basis of $\left.\mathcal{M}\times \mathbb{R}^{n+1}\right|_{U}\simeq \mathbb{R}^{n+1}$, {\it i\,.e\,.,} $\xi_{i}$, to a linear combination of a basis of $\left.T\mathcal{M}\right|_{U}\simeq T_{p}\mathcal{M}$ $(p\in U)$, where ``$\simeq$'' denotes the isomorphic relation between two objects. The basis of $T_{p}\mathcal{M}$ can be generically taken arbitrarily, but we use the standard coordinate basis, {\it i\,.e\,.,} $\partial_{\mu}$, to a chart of an atlas of $\mathcal{M}$. Explicitly, on an open set $U$, we can express this relation as follows:
\begin{equation}
e_{i}:={\bf e}(\xi_{i})= e_{i}{}^{\mu}\partial_{\mu}\,.
\label{}
\end{equation}
The frame field $\bf{e}$ has its inverse in a {\it local} region of $\mathcal{M}$, although it is not true in a {\it global} region in general. The construction of $\bf{e}$ leads to the fact that if we take a local region as an open set of the open cover of $\mathcal{M}$ then $\bf{e}$ always has its inverse under the restriction to the local region. Let us take the open set $U$ as such local region. Then we can define the inverse map of $\bf{e}$, {\it i\,.e\,.,} ${\bf e}^{-1}:\left.T\mathcal{M}\right|_{U}\rightarrow\left.\mathcal{M}\times \mathbb{R}^{n+1}\right|_{U}$, and the explicit formula as follows:
\begin{equation}
e^{i}:=({\bf e}^{-1})^{*}(\xi^{i})=e^{i}{}_{\mu}dx^{\mu}\,,
\label{}
\end{equation}
where we denote $({\bf e}^{-1})^{*}$ as the pull-back operator of ${\bf e}^{-1}$. This inverse ${\bf e}^{-1}$ is called as co-frame field of $\bf{e}$ on the open region $U$. The dual structure derives the relation between $e_{i}{}^{\mu}$ and $e^{i}{}_{\mu}$: $e^{\mu}{}_{i}e^{i}{}_{\nu}=\delta^{\mu}_{\nu}$ and $e_{i}{}^{\mu}e^{j}{}_{\mu}=\delta^{j}_{i}$. In terms of these quantities, the components of the metric tensor $g=g_{\mu\nu}dx^{\mu}\otimes dx^{\nu}$ on $\mathcal{M}$ is related to that on $M\times \mathbb{R}^{1,n}$, {\it i\,.e\,.,} $g=g_{ij}\xi^{i}\otimes\xi^{j}$, as follows:
\begin{equation}
e_{i}{}^{\mu}e_{j}{}^{\nu}g_{\mu\nu}=g_{ij}
\label{vielbein to metric}
\end{equation}
or, if $\bf{e}$ is restricted to the local region in which it has its inverse, we also have 
\begin{equation}
g_{\mu\nu}=e^{i}{}_{\mu}e^{j}{}_{\nu}g_{ij}\,.
\label{metric to vielbein}
\end{equation}
That is, the invertibility of the frame field connects the metric tensor on the space-time to that on the internal space in a one-to-one manner. 

In order to introduce the concept of covariant derivative into space-time and internal space, we define the connection as usual. For the spacetime, the affine connection is denoted as $\tilde{\Gamma}^{\rho}{}_{\mu\nu}$. For the internal space, we introduce the spin connection as follows:
\begin{equation}
\mathcal{D}_{\mu}e_{i}:=\omega^{j}{}_{i\mu}e_{j}
\label{}
\end{equation}
where we used the same notation to the affine connection~\cite{Nakahara2003}. In particular, since $\left.\mathcal{M}\times \mathbb{R}^{n+1}\right|_{U}\simeq \left.T\mathcal{M}\right|_{U}$ holds in the local region $U$, we can add $\tilde{\Gamma}$ and $\omega$ together, and we get the covariant derivative of co-frame field components as follows:
\begin{equation}
\mathcal{D}_{\mu}e^{i}{}_{\nu}=\partial_{\mu}e^{i}{}_{\nu}-\tilde{\Gamma}^{\rho}{}_{\mu\nu}e^{i}{}_{\rho}+\omega^{i}{}_{j\mu}e^{j}{}_{\nu}\,.
\label{local covariant derivative formula}
\end{equation}
This relation plays a crucial role to give the attribute of an internal gauge symmetry to gravity theories at each space-time point. In fact, for a Lie group $G$, the co-frame field transformation $e^{i}{}_{\mu}\rightarrow e'^{i}{}_{\mu}=\Lambda^{i}{}_{j}e^{j}{}_{\mu}$ $(\Lambda^{i}{}_{j}\in G)$ leads to 
\begin{equation}
\mathcal{D}_{\mu}e'^{i}{}_{\nu}=\Lambda^{i}{}_{j}\mathcal{D}_{\mu}e^{j}{}_{\nu}
\label{covariant derivative operator}
\end{equation}
where the spin connection transforms as follows:
\begin{equation}
\omega^{i}{}_{j\mu}\rightarrow \omega'^{i}{}_{j\mu}=(\Lambda^{-1})^{i}_{\ k}\partial_{\mu}\Lambda^{k}{}_{j}+(\Lambda^{-1})^{i}{}_{k}\Lambda^{l}{}_{j}\omega^{k}{}_{l\mu}\,.
\label{gauge transformation}
\end{equation}
This is nothing but the gauge transformation law of the spin connection in the usual manner. Remark that the same arguments hold even for the frame field components as long as we consider the local region in which the frame field is invertible. 

Finally, notice that we have an important relation between the affine connection and the spin connection, {\it i\,.e\,.}, the {\it ``frame field (or vielbein) postulate"}:
\begin{equation}
\mathcal{D}_{\mu}e^{i}{}_{\nu}=0\,.
\label{vielbein postulate}
\end{equation}
This relation always holds as an identity in the local region which makes the addition of the affine connection and the spin connection well-defined~\cite{Carroll1997}. The postulate also allows to express the affine connection in terms of the co-frame field components and the spin connection as follows:
\begin{equation}
\tilde{\Gamma}^{\rho}{}_{\mu\nu}=e_{i}{}^{\rho}\partial_{\mu}e^{i}{}_{\nu}+\omega^{i}{}_{j\mu}e_{i}{}^{\rho}e^{j}{}_{\nu}
\label{relation btw affine and spin}
\end{equation}
by using the derivative formula Eq~$(\ref{local covariant derivative formula})$. This formula does not depend on the gauges by virtue of the relation Eqs~$(\ref{covariant derivative operator})$ and~$(\ref{gauge transformation})$. 

Armed with Eqs~$(\ref{metric to vielbein})$ and~$(\ref{relation btw affine and spin})$, a gravity theory is reformulated in terms of the (co-)frame field and the spin connection. Let us consider the Einstein-Palatini action:
\begin{equation}
S_{\rm  EP}[g_{\mu\nu},\tilde{\Gamma}^{\rho}{}_{\mu\nu}]:=\int_{\mathcal{M}}d^{n+1}x\sqrt{-g}\ \tilde{R}
\label{EP action}
\end{equation}
where $g$ is the determinant of the metric tensor $g_{\mu\nu}$ and $\tilde{R}$ is general the Ricci scalar.
In this action, gravity is described by the independent variables: $g_{\mu\nu}$ and $\tilde{\Gamma}^{\rho}{}_{\mu\nu}$. Utilizing Eqs~$(\ref{metric to vielbein})$ and~$(\ref{relation btw affine and spin})$, the variables are replaced by the co-frame fields and the spin connection, as follows:
\begin{equation}
\hat{S}_{\rm  EP}[e^{i}{}_{\mu},\omega^{i}{}_{j\mu}]:=\int_{\mathcal{M}}d^{n+1}x\ {\rm det}({\bf e}^{-1})\ \hat{R}
\label{gauge theorized EP action}
\end{equation}
where ${\rm det}({\bf e}^{-1})$ is the determinant of the co-frame field components~\footnote{We can identify the internal space index ``$i$'' and the space-time index ``$\mu$'' in a local region by virtue of $\left.M\times \mathbb{R}^{1,n}\right|_{U}\simeq \left.TM\right|_{U}$.} and the hat ``\textasciicircum'' denotes the
quantities that are described by the co-frame fields and the spin connection. The gauge group is set as $G=T^{1,n}\rtimes SO(1,n)$, where $T^{1,n}$ denotes the translation group in a $(n+1)$-dimensional Minkowskian spacetime. This action is also called the ``first-order formulation of GR". The spin connection for this internal symmetry is called the Levi-Civita (or Lorentz) connection \cite{Baez1994,Nakahara2003}. The theory has now the gauge symmetry of $T^{1,n}\rtimes SO(1,n)$ at each space-time point. 
Remark that the procedure is applicable to any theory of gravity constructed from gauge invariants. 

\subsection{\label{sec:02:02}Metric-affine gauge theory of gravity}
GR describes gravity in terms of geometrical quantities of (pseudo-)Riemannian geometry based on the equivalence principle. In this geometry, only the Riemannian curvature tensor plays the main role to describe gravity. That is, it assumes that the torsion and the non-metricity vanishes in advance. However, there are other possibilities to take these two geometrical quantities into account. This generalized geometry is called metric-affine geometry~\cite{Hehl1995}.

First of all, we introduce the fundamental quantities to formulate the geometry. The covariant derivative is defined as follows:
\begin{equation}
\tilde{\nabla}_{\mu}A^{\nu}=\partial_{\mu}A^{\nu}+\tilde{\Gamma}^{\nu}{}_{\rho\mu}A^{\rho}\,,
\label{covariant derivative in MAG}
\end{equation}
where $\tilde{\Gamma}^{\nu}{}_{\rho\mu}$ denotes the affine connection and $A^{\nu}$ are the contra-variant vector components. The important point here is that in the above definition, it does generically not allow to commute with the lower two indices of the affine connection: the order has a specific meaning. That is, it gives the torsion tensor of the geometry:
\begin{equation}
T^{\rho}_{\ \ \mu\nu}=\tilde{\Gamma}^{\rho}{}_{\mu\nu}-\tilde{\Gamma}^{\rho}{}_{\nu\mu}:=2\tilde{\Gamma}^{\rho}{}_{[\mu\nu]}\,.
\label{torsion tensor}
\end{equation}
In order to manipulate the indices, the covariant derivative of the metric tensor is important; if it vanishes then the metric tensor can freely move inside and outside of the covariant derivative, but if it is not the case then this manipulation does not hold. The non-metricity tensor of the geometry governs this manipulation:
\begin{equation}
Q_{\rho\mu\nu}:=\tilde{\nabla}_{\rho}g_{\mu\nu}\,.
\label{non-metricity tensor}
\end{equation}
Using these quantities, the affine connection is decomposed into as follows:
\begin{equation}
\tilde{\Gamma}^{\rho}{}_{\mu\nu}=\lc{\Gamma}^{\rho}{}_{\mu\nu}+K^{\rho}{}_{\mu\nu}+L^{\rho}{}_{\mu\nu}
\label{connection decomposition}
\end{equation}
where $\lc{\Gamma}^{\rho}{}_{\mu\nu}$ is the Christoffel symbols, $K^{\rho}{}_{\mu\nu}$ is the contortion tensor:
\begin{equation}
K^{\rho}{}_{\mu\nu}=\frac{1}{2}T^{\rho}{}_{\mu\nu}+T_{(\mu\ \ \nu)}^{\ \ \rho\ }
\label{con-torsion tensor}
\end{equation}
and $L^{\rho}_{\ \mu\nu}$ is the disformation tensor:
\begin{equation}
L^{\rho}{}_{\mu\nu}=\frac{1}{2}Q^{\rho}{}_{\mu\nu}-Q_{(\mu\ \nu)}^{\ \ \rho\ }\,.
\label{disformation tensor}
\end{equation}
The curvature tensor is introduced in terms of the affine connection $\tilde\Gamma^{\rho}{}_{\mu\nu}$ as usual:
\begin{equation}
\tilde{R}^{\sigma}{}_{\mu\nu\rho}=2\partial_{[\nu}\tilde{\Gamma}^{\sigma}{}_{\rho]\mu}+2\tilde{\Gamma}^{\sigma}{}_{[\nu|\lambda|}\tilde{\Gamma}^{\lambda}{}_{\rho]\mu}\,.
\label{curvature tensor in MAG}
\end{equation}
Here, remark again that the position of the indices in the affine connection is crucial, unlike the ordinary Riemannian curvature tensor. If the affine connection is decomposed as $\tilde{\Gamma}^{\rho}{}_{\mu\nu}=\lc{\Gamma}^{\rho}{}_{\mu\nu}+N^{\rho}{}_{\mu\nu}$ for a distortion tensor $N^{\rho}{}_{\mu\nu}$, 
a straightforward computation derives the following formula:
\begin{equation}
\tilde{R}^{\sigma}{}_{\mu\nu\rho}=\lc{R}^{\sigma}{}_{\mu\nu\rho}+2\lc{\nabla}_{[\nu}N^{\sigma}{}_{\rho]\mu}+2N^{\sigma}{}_{[\nu|\lambda|}N^{\lambda}{}_{\rho]\mu}\,,
\label{formula of curvature tensor in MAG}
\end{equation}
where $\lc{R}^{\sigma}{}_{\mu\nu\rho}$ is the (pseudo-)Riemannian curvature tensor and $\lc{\nabla}_{\nu}$ denotes the covariant derivative defined by the Christoffel symbols.

Using the curvature tensor Eq~(\ref{curvature tensor in MAG}), the Einstein-Palatini action Eq~$(\ref{EP action})$ is now described by the metric-affine geometry and there are different types of geometry, depending on whether or not the torsion, the non-metricity, and the curvature tensor vanishes, respectively. As a special case, imposing conditions that the torsion and the non-metricity tensor vanish, the Einstein-Hilbert action~\cite{Hilbert1915,Einstein1916} is recovered:
\begin{equation}
S_{\rm EH}[g_{\mu\nu}]:=\int_{\mathcal{M}}d^{n+1}x\sqrt{-g}\ \lc{R}
\label{EH action}\,.
\end{equation}
One can also construct more general theories in this framework belonging to the general linear gauge group: $T^{n+1}\rtimes GL(n+1,\mathbb{R})$~\cite{Hehl1995}, where $T^{n+1}$ denotes the translation group in a $(n+1)$-dimensional Euclidean space. Theories constructed from scalars that are invariant under that group are called {\it ``metric-affine gauge theory of gravity"}.

\subsection{\label{sec:02:03}Teleparallelism and the geometrical trinity of gravity}
The metric-affine gauge theory of gravity has intriguing branches that are equivalent to GR up to surface terms. In order to derive these branches, the so-called {\it ``teleparallel condition" (or ``teleparallelism")} is imposed as follows:~\footnote{Note that quantities without any symbol on top refer to Teleparallel ones.}
\begin{equation}
\tilde{R}^{\sigma}{}_{\mu\nu\rho}:=0=R^{\sigma}{}_{\mu\nu\rho}\,.
\label{teleparallelism}
\end{equation}
Under this condition, the affine connection can be resolved at least in a local region as follows:
\begin{equation}
\Gamma^{\rho}{}_{\mu\nu}=e_{i}{}^{\rho}\partial_{\mu}e^{i}{}_{\nu}\,.
\label{affine connection in Weitzenboch gauge}
\end{equation}
One can check this statement by substituting Eq~(\ref{affine connection in Weitzenboch gauge}) into Eq~(\ref{teleparallelism}). Note that, in a local region, for any vector bundles, the so-called standard flat connection, that is $\omega^{i}{}_{j\mu}=0$, exists~\cite{Baez1994,Nakahara2003}, and the Eq~(\ref{relation btw affine and spin}) implies the existence of the solution. This condition is sometimes called the ``{\it Weitzenb\"{o}ch gauge}''~\cite{Weitzenboh1923,Blagojevic2020,Jimenez2022}.

In addition to the teleparallel condition, since the metric-affine gauge theory of gravity has three independent geometrical quantities: curvature, torsion, and non-metricity, it is possible to impose further conditions. The imposition of vanishing non-metricity leads to the so-called {\it ``Teleparallel Equivalent to GR" (TEGR)}~\cite{Jimenez2019} and the affine connection is provided by the solution of the following equation:~\footnote{
Eq~(\ref{eq for TEGR connection}) has a solution as follows: $g_{\mu\nu}=e^{i}{}_{\mu}e^{j}{}_{\nu}c_{ij}$, where $c_{ij}$ is an arbitrary non-singular symmetric constant tensor. This solution is a special case of Eq~(\ref{metric to vielbein}). Therefore, if we chose the gauge for $g_{ij}$ as a constant tensor $c_{ij}$ then the condition of vanishing non-metricity is satisfied. 
}
\begin{equation}
2e_{i}{}^{\rho}\partial_{\beta}e^{i}{}_{(\mu}g_{\nu)\rho}=\partial_{\beta}g_{\mu\nu}\,.
\label{eq for TEGR connection}
\end{equation}
Using the formula Eqs~(\ref{formula of curvature tensor in MAG}) and~(\ref{teleparallelism}), we can show the following relation:
\begin{equation}
\tilde{R}=\lc{R}+T-\lc{\nabla}_{\mu}T^{\mu}=0\,,
\label{Ricci scalar in TEGR}
\end{equation}
where 
\begin{equation}
T:=-\frac{1}{4}T_{\alpha\mu\nu}T^{\alpha\mu\nu}-\frac{1}{2}T_{\alpha\mu\nu}T^{\mu\alpha\nu}+T^{\alpha}T_{\alpha}\,,
\label{}
\end{equation}
and $T_{\alpha}:=T^{\mu}{}_{\mu\alpha}$. Neglecting the boundary term, therefore, the Einstein-Palatini action Eq~(\ref{EP action}) leads to the TEGR action:
\begin{equation}
S_{\rm TEGR}[g_{\mu\nu}]:=-\int_{\mathcal{M}}d^{n+1}x\sqrt{-g}\ T\,,
\label{TEGR action}
\end{equation}
and this action is equivalent to the Einstein-Hilbert action Eq~(\ref{EH action})~ excepting the geometry and neglecting boundary terms. Applying the procedure in Sec.~\ref{sec:02:01}, $\sqrt{-g}$ and $T$ are just replaced by ${\rm det}(e^{-1})$ and $\hat{T}$, respectively, and the variables describing the system are the (co-)frame field

In the same manner, the imposition of vanishing torsion leads to the so-called {\it ``Symmetric Teleparallel Equivalent to GR" (STEGR)}~\cite{Jimenez2019} and the affine connection {\it is solved} as follows
:
\begin{equation}
\Gamma^{\rho}{}_{\mu\nu}=\frac{\partial x^{\rho}}{\partial \zeta^{i}}\partial_{\mu}\partial_{\nu}\zeta^{i}\,,
\label{connection in STEGR}
\end{equation}
where $\zeta^{i}$ are arbitrary functions~\footnote{These functions are none others than the so-called St\"{u}ckelberg fields \cite{Jimenez2022}.} defined on a local region $\left.M\times \mathbb{R}^{1,n}\right|_{U}\simeq \left.TM\right|_{U}$.~\footnote{See footnote 1. This local property plays an essential role to formulate the coincident GR.
}
Using the formula Eq~(\ref{curvature tensor in MAG}), we get the following equation:
\begin{equation}
\tilde{R}=\lc{R}-Q+
\lc{\nabla}_{\mu}(Q^{\mu}-\tilde{Q}^{\mu})=0\,,
\label{Ricci scalar in STEGR}
\end{equation}
where
\begin{equation}
Q:=-\frac{1}{4}Q_{\mu\nu\alpha}Q^{\mu\nu\alpha}+\frac{1}{2}Q_{\mu\nu\alpha}Q^{\nu\mu\alpha}+\frac{1}{4}Q_{\alpha}Q^{\alpha}-\frac{1}{2}Q_{\alpha}\tilde{Q}^{\alpha}\,,
\label{}
\end{equation}
$Q_{\alpha}:=Q_{\alpha\mu}{}{}^{\mu}$, and $\tilde{Q}_{\alpha}:=Q^{\mu}{}_{\mu\alpha}$.
The Einstein-Palatini action Eq~(\ref{EP action}) leads to the STEGR action as follows:
\begin{equation}
S_{\rm  STEGR}[g_{\mu\nu}]:=\int_{\mathcal{M}}d^{n+1}x\sqrt{-g}\ Q\,.
\label{STEGR action}
\end{equation}
Applying the procedure in Sec.~\ref{sec:02:01}, $\sqrt{-g}$ and $Q$ are just replaced by ${\rm det}({\bf e}^{-1})$ and $\hat{Q}$, respectively, and the variables describing the system are the (co-)frame field. 

So far we obtain three specific gravity theories: GR, TEGR, and STEGR. These three gravity theories are equivalent up to boundary terms and called the {\it ``geometrical trinity of gravity"}~\cite{Jimenez2019}. In this paper, we focus on the STEGR branch and its extensions. 

\subsection{\label{sec:02:04}Coincident GR}
In the STEGR branch, the connection is easily solved as in Eq~(\ref{connection in STEGR}). Again, noticing that the local relation of $\left.\mathcal{M}\times \mathbb{R}^{1,n}\right|_{U}\simeq \left.T\mathcal{M}\right|_{U}$, we can impose further gauge condition on STEGR. Since the functions $\zeta^{i}$ are defined on the local region $U$, it can be expressed by the coordinates system, {\it i\,.e\,.,} $x^{\mu}$, for $U\subset \mathcal{M}$: $\zeta^{i}=\zeta^{i}(x)$. Therefore, in this local region, $\zeta^{i}$ are expanded in terms of $x^{\mu}$ up to first order terms as follows:
\begin{equation}
\zeta^{i}=M^{i}{}_{\mu}x^{\mu}+A^{i}\,,
\label{coincident gauge}
\end{equation}
where $M^{i}{}_{\mu}\in GL(n+1,\mathbb{R})$~\footnote{This group is not a Lie group: a global symmetry to the internal space.} and $A^{i}$ are arbitrary constant $(n+1)$-vector components. This is just an affine transformation in the internal space. Then the connection given in Eq~(\ref{connection in STEGR}) becomes as follows:
\begin{equation}
\Gamma^{\rho}{}_{\mu\nu}=0\,.
\label{affine connection in CG}
\end{equation}
Under imposing this new gauge condition, or the {\it ``coincident gauge condition"}, {\it i\,.e\,.,} Eq~(\ref{coincident gauge}), it reveals that STEGR has a more specific branch. This branch is called {\it ``Coincident GR" (CGR)}~\cite{Jimenez2018}.

The equation Eq~(\ref{affine connection in CG}) implies the equivalence to GR without boundary terms. That is, the decomposition Eq~(\ref{connection decomposition}) with Eq~(\ref{affine connection in CG}) leads to the following relation:
\begin{equation}
L^{\rho}{}_{\mu\nu}=-\lc{\Gamma}{^{\rho}{}_{\mu\nu}}\,.
\label{disformation in CG gauge}
\end{equation}
Neglecting boundary terms, therefore, Eq~(\ref{STEGR action}) under the coincident gauge derives the following action~\footnote{
Remark that this action was first derived by A. Einstein in 1916~\cite{Einstein1916}, which was based on the well-posedness of the variational principle under the Dirichlet boundary conditions, although there are some controversies even in nowadays~\cite{Keisuke2023,Kyosuke2023}. Therefore, it is a revisiting of his work from the viewpoint of a modern perspective, that is, the gauge theory of gravity. 
}:
\begin{equation}
\begin{split}
S_{\rm CGR}=&\int_{\mathcal{M}}d^{n+1}x\sqrt{-g}\ 2L^{\rho}{}_{[\rho|\lambda|}L^{\lambda}{}_{\nu]\mu}=\int_{\mathcal{M}}d^{n+1}x\sqrt{-g}\ 2\lc{\Gamma}^{\rho}{}_{\lambda[\rho}\lc{\Gamma}{^{\lambda}{}_{\mu]\nu}}\,.
\end{split}
\label{CGR-action}
\end{equation}
This is none other than the Einstein-Hilbert action without the boundary term~\cite{Einstein1916,Padmanabhan2006}. From this perspective, we would expect that CGR is equivalent to GR as a constraint system; the Poisson Bracket algebra (PB-algebra) and the propagating Degrees of Freedom (pDoF) would be coincident. 

\section{\label{sec:03}Hamiltonian analysis of GR and \texorpdfstring{$f(\mathring{R})$}{R}-gravity}
In order to apply the Dirac-Bergmann analysis (See Appendix~\ref{sec:03:01} in detail) to field theories, it needs careful manipulations of boundary terms. In this section, we reveal that Gibbons-York-Hawking type boundary terms well-known in GR~\cite{York1972,GibbonsHawking1977,York1986,HawkingHorowitz1996} can be neglected without any change in the symplectic structure of a given system when performing the analysis. We also provide a prescription to circumvent the problematic situation concerning the PDEs of Lagrange multipliers which is mentioned in a series of works~\cite{Sundermeyer:1982,Blagojevic2020,Fabio2023}. Finally, we revisit the analysis of GR and $f(\lc{R})$-gravity to demonstrate how to work the statements declared in this section.
\subsection{\label{sec:03:02}A role of surface terms in 
Dirac-Bergmann analysis}
Symplectic structure plays the most fundamental role in analytical mechanics since once the structure and a total Hamiltonian are given, then, the dynamics are uniquely determined. This statement is verified from the following two facts; (i) The definition of the Poisson bracket: $\{f,g\}:=\Omega(X_{f},X_{g})$, where $X_{f}$ and $X_{g}$ are the Hamiltonian vector fields with respect to some functions $f$ and $g$, respectively, and $\Omega$ is a symplectic form of the system; (ii) The time development of a quantity $F$ of the system is, of course, given by $\dot{F}=\{F,H_{T}\}$. Therefore, under a given total Hamiltonian, the symplectic structure governs everything in the system.

To clarify a relation between the symplectic structure and surface terms, let us consider the symplectic potential: $\omega:=p_{i}d q^{i}+dW$, where $W=W(q^{i})$ is an arbitrary function. This quantity is just the integral of the symplectic form $\Omega$ and has arbitrariness of $W$. In fact, one can easily verify that $d\omega=\Omega$. Then, notice that the first terms of $\omega$, $p_{i}\delta q^{i}$, are none other than the surface term of the first variation of the Lagrangian in Eq~(\ref{FVofPL}). This relation, therefore, implies that the Lagrangian has also arbitrariness of surface terms: $L\rightarrow L'=L+dW/dt$ for the common $W$, and the first-order variation of $L'$ becomes $\delta L':=\left[{\rm{EoM}}\right]_{i}\delta q^{i}+d(p'_{i}\delta q^{i})/dt$, where $p'_{i}:=p_{i}+\partial W/\partial q^{i}$. Since $\omega':=p'_{i}\delta q^{i}=\omega$, all arguments are consistent, and $\Omega$ does not depend on the difference of symplectic potentials. Therefore, we conclude an important proposition; {\it Surface terms do change canonical momentum variables but do not change the symplectic structure}. 

So far, we consider first-order derivative systems, but when treating gravity theories including GR, we need a theory of degenerate second-order derivative systems from the perspective of the well-posedness of the variational principle, and it is inevitable to intervene surface terms. To clarify this statement, let us consider the following Lagrangian:
\begin{equation}
L=L(\ddot{q}^{i},\dot{q}^{i},q^{i}),
\label{PL2nd}
\end{equation}
where $i\in\{1,2,\cdots,n\}$. The first-order variation of this Lagrangian is calculated as follows:
\begin{equation}
\delta L=\left[\frac{\partial L}{\partial q^{i}}-\frac{d}{dt}\frac{\partial L}{\partial \dot{q}^{i}}+\frac{d^{2}}{dt^{2}}\frac{\partial L}{\partial \ddot{q}^{i}}\right]\delta q^{i}+\frac{d}{dt}\left[\left(\frac{\partial L}{\partial \dot{q}^{i}}-\frac{d}{dt}\frac{\partial L}{\partial \ddot{q}^{i}}\right)\delta q^{i}+\left(\frac{\partial L}{\partial \ddot{q}^{i}}\right)\delta\dot{q}^{i}\right]:=\left[{\rm EoM}\right]_{i}\delta q^{i}+\frac{d}{dt}\left[p^{(1)}_{i}\delta q^{i}+p^{(2)}_{i}\delta\dot{q}^{i}\right]\,.
\label{FVofPL2nd}
\end{equation}
$p^{(1)}_{i}$ and $p^{(2)}_{i}$ are canonical momentum variables of the system. The Hessian matrix is defined as follows:
\begin{equation}
K^{(2)}_{ij}:=\frac{\partial p^{(2)}_{i}}{\partial \ddot{q}^{i}}=\frac{\partial^{2}L}{\partial\ddot{q}^{i}\partial\ddot{q}^{j}}\,,\ \ \ K^{(1)}_{ij}:=\frac{\partial p^{(1)}_{i}}{\partial\dot{q}^{j}}=\frac{\partial^{2}L}{\partial\dot{q}^{i}\partial\dot{q}^{j}}-\frac{d}{dt}\frac{\partial^{2} L}{\partial\ddot{q}^{i}\partial\dot{q}^{j}}\,.
\label{HesseMatrix2nd}
\end{equation}
Let us assume that the ranks of these matrices are $0$ and $n-r^{(1)}$, respectively.~\footnote{
When the rank of the first Hesse matrix does not vanish, it may give rise to third- and/or fourth-order derivative equations of motion. Under the imposition of appropriate conditions, such systems can also describe dynamics without the Ostrogradski instability~\cite{Ostrogradsky1850,Woodard2015} just like DHOST~\cite{Langlois2016,Crisostomi2016,Achour2016} but these topics are out of scope of the current paper.
}~\footnote{
Precisely speaking, in order to make the equations of motion up to second-order time derivative, the rank of the matrix $E_{ij}:=\partial p^{(2)}_{i}/\partial \dot{q}^{j}-\partial p^{(2)}_{j}/\partial \dot{q}^{i}$ have also to be zero.
}
Then the number of $n+r^{(1)}$ primary constraints appears. These constraints are derived in the same manner to the first-order theory as follows: $\phi^{(1)}_{\alpha^{(2)}}:=p^{(2)}_{\alpha^{(2)}}-f_{\alpha^{(2)}}(q^{i}_{(1)},q^{i}_{(2)},p^{(1)}_{i},p^{(2)}_{i}):\approx0$ $(\alpha^{(2)}\in\{1,2,\cdots,n\})$ and $\phi^{(1)}_{n+\alpha^{(1)}}:=p^{(1)}_{\alpha^{(1)}}-g_{\alpha^{(1)}}(q^{i}_{(1)},q^{i}_{(2)},p^{(1)}_{i},p^{(2)}_{i}):\approx0$ $(\alpha^{(1)}\in\{1,2,\cdots,r^{(1)}\})$, where $q^{i}_{(1)}:=q^{i}, q^{i}_{(2)}:=\dot{q}^{i}$~\cite{Sugano1989,Sugano1993,Pons1989}. Let us denote the phase subspace which is restricted by these primary constraints as $\mathfrak{C}^{(1)}$. Since the variational principle is not well-posed until appropriate boundary conditions are imposed, the equations of motion cannot be derived in a consistent manner to the degeneracy of the system. This indicates that it needs careful consideration for the application of the Dirac-Bergmann analysis. 

The symplectic structure of the system is given as follows: $\Omega=dq^{i}_{(2)}\wedge dp^{(2)}_{i}+dq^{i}_{(1)}\wedge dp^{(1)}_{i}$. Therefore, the symplectic potential becomes $\omega=p^{(2)}_{i}dq^{i}_{(2)}+p^{(1)}_{i}dq^{i}_{(1)}+dW$, where $W=W(q^{i}_{(1)},q^{i}_{(2)})$ is arbitrary function. The same consideration to the first-order theory leads to new canonical momentum variables: $p'^{(2)}_{i}:=p^{(2)}_{i}+\partial W/\partial q^{i}_{(2)}$ and $p'^{(1)}_{i}:=p^{(1)}_{i}+\partial W/\partial q^{i}_{(1)}$ without any changing the symplectic structure. Since the Hessian matrices are not changed by this manipulation, $K'^{(2)}_{ij}=\partial^{2}L'/\partial\dot{q}^{i}_{(2)}\partial\dot{q}^{j}_{(2)}=K^{(2)}_{ij}$, where $L\rightarrow L'=L+dW/dt$, so does the rank of the matrix. $K^{(1)}$ has the same property. These properties imply that it exists a surface term $W$ in $\mathfrak{C}^{(1)}$ such that $\phi'^{(1)}_{\alpha'^{(2)}}:=p'^{(2)}_{\alpha'^{(2)}}:\approx0$ $(\alpha'^{(2)}\in\{1,2,\cdots,r'^{(2)}\leq n\})$ and $\phi'^{(1)}_{r'^{(2)}+\alpha'^{(1)}}:=p'^{(1)}_{\alpha'^{(1)}}:\approx0$ $(\alpha'^{(1)}\in\{1,2,\cdots,r'^{(1)}\leq r^{(1)}\})$. Armed with these facts, to make the variational principle well-posed, it is necessary to impose boundary conditions that are consistent with the primary constraints: $\delta q^{a'^{(2)}}_{(2)}=0$ $(a'^{(2)}\in\{1,2,\cdots,n-r'^{(2)}\})$ and $\delta q^{a'^{(1)}}_{(1)}=0$ $(a'^{(1)}\in\{1,2,\cdots,n-r'^{(1)}\})$. In particular, for the case of $r'^{(2)}=n$,~\footnote{
The case can be realised if the rank of $E_{ij}$ is zero with appropriate boundary (counter) terms like Gibbons-York-Hawking term~\cite{York1972,GibbonsHawking1977,York1986,HawkingHorowitz1996,Keisuke2023,Kyosuke2023}. Remark that the manipulation does not change the symplectic structure, {\it i\,.e\,.} the time evolutin of the system, as mentioned in the main manuscript.
}
the boundary conditions become $\delta q^{a'^{(1)}}_{(1)}=0$, and then the absence of the Ostrogradski instability is guaranteed~\cite{Ostrogradsky1850,Woodard2015,Kyosuke2023}. Then the Dirac-Bergmann analysis becomes applicable.~\footnote{Note that the surface term $W$ is none other than the so-called counter-term that appears in higher-order derivative systems just like Gibbons-York-Hawking counter-terms in GR~\cite{York1972,GibbonsHawking1977,York1986,HawkingHorowitz1996}. Recently, a different sort of counter-term was proposed~\cite{Keisuke2023}, which is based on the requirement of the imposition of boundary conditions for the well-posed variational principle that originated from the consistency with the {\it full} result of the Dirac-Bergmann analysis. That is the new sort of counter-term demands consistency with $\mathfrak{C}^{(K)}$ rather than $\mathfrak{C}^{(1)}$. 
}
The essential message here is that the well-posedness does not affect the symplectic structure. Therefore, we conclude an important proposition; {\it The Dirac-Bergmann analysis is applicable also in second-order derivative systems without depending on the well-posedness of the variational principle}. This statement means that, when we use the Dirac-Bergmann analysis, all surface terms can be neglected freely.

There is a remark on when applying the Dirac-Bergmann analysis to field theories. Departing from theories of point particle systems, field theories generically have spatial boundary terms. Since spatial boundary terms have no relation to the symplectic structure, there is no concern with the problem of the well-posedness of the variational principle. It implies that spatial boundary conditions can be taken arbitrarily. Precisely speaking, spatial boundary terms are concerned only with the continuum limit of boundaries in field theories~\cite{Jabbari1999}; it does not affect the dynamics (time evolution). Recently, the authors in~\cite{Fabio2023} stated that the existence of such spatial boundary terms might break the Dirac-Bergmann analysis but it is not the case when we hold the following prescription, or more generically speaking when taking into account that the spatial boundary terms can always be neglected by imposing appropriate spatial boundary conditions in the variational principle and it never affects the dynamics (time evolution).

This fact leads to a convenient prescription when computing PB-algebras explicitly since the PB-algebras are defined on a hypersurface that has the common spatial boundary as that of the variational principle in the ADM-foliation~\cite{ADM1959,ADM1960};\newline\newline
{\it For some field $A(x)$ on a $(n+1)$-dimensional spacetime, the term ``$\sqrt{h}A(x)\partial^{(x)}_{I}\delta^{(n)}(\vec{x}-\vec{y})$'', where $I$ runs from $1$ to the dimension of the hypersurface $n$, in PB-algebras can be neglected by setting properly spatial boundary conditions of $A(x)$ in the variational principle, where $h$ is the determinant of the metric of a $n$-dimensional hypersurface.}\newline\newline
That is, the problematic term ``$\sqrt{h}A(x)\partial^{(x)}_{I}\delta^{(n)}(\vec{x}-\vec{y})$'' in the PB-algebras can be neglected since integrating by parts it on $\Sigma_{t}$ and integrating all terms again on $\Sigma_{t}$ then we obtain only spatial boundary terms on $\partial\Sigma_{t}$; these terms can be vanished by the imposition of appropriate spatial boundary conditions if it is necessary. In this paper, we assume these propositions and the prescription that are introduced above and then this analysis works, and then, we can derive the final result of the pDoF. However, we emphasize that this is our argument against the paper~\cite{Fabio2023} and this is of course a debatable point to consider.

\subsection{\label{sec:03:03}Revisiting to Hamiltonian analysis of GR and \texorpdfstring{$f(\mathring{R})$}{R}-gravity}
Based on the previous subsection, let us count the pDoF of GR and $f(\lc{R})$-gravity as examples. First of all, we review the Dirac-Bergmann analysis of GR. We use notations and set-ups that are fixed in this subsection throughout the subsequent sections of the paper. 

Let $\mathcal{M}$ and $g_{\mu\nu}$ be $(n+1)$-dimensional spacetime and its metric tensor, respectively. Then ADM-foliation~\cite{Baez1994,ADM1959,ADM1960} of $\mathcal{M}$ is a diffeomorphism $\sigma: \mathcal{M}\rightarrow\mathbb{R}\times \mathcal{S}^{n}$ such that it decomposes $\mathcal{M}$ as a disjoint union of hypersurfaces $\Sigma_{t}:=\{p\in\mathcal{M}|\sigma^{*}\tau(p):=t\}$, which is deffeomorphic to $\{t\}\times\mathcal{S}^{n}$, {\it i\,.e\,.,} $\mathcal{M}=\sqcup_{t\in \mathcal{I}}\Sigma_{t}$, where $\mathcal{S}^{n}$ denotes a $n$-dimensional hypersurface, $\mathcal{I}$ is a time-interval of $\mathcal{M}$, $t$ is a time-coordinate of $\mathcal{M}$: $t\in \mathcal{I}$, $\tau$ is a time-coordinate of $\mathbb{R}\times \mathcal{S}^{n}$, and $\sigma^{*}$ is the pullback operator of the diffeomorphism $\sigma$. Then the metric of the spacetime is decomposed as follows:
\begin{equation}
ds^{2}=
-N^{2}dt^{2}+h_{IJ}\left(dx^{I}+N^{I}dx^{I}\right)\left(dx^{J}+N^{I}dx^{J}\right)\,,
\label{ADM-metric}
\end{equation}
where $I,J\in\{1,2,\cdots,n\}$, $N:=N(x)$ and $N^{I}:=N^{I}(x)$ are a lapse function and a shift vector, respectively. Then the normal vector $n^{\mu}$ to a leaf $\Sigma_{t}$ is given by $n^{\mu}=
-N^{-1}(-1,N^{I})$. $n^{\mu}$ satisfies the normalization condition: $n^{\mu}n_{\mu}=
-1$. 

To reconstruct GR on the ADM-foliated spacetime $\mathbb{R}\times \mathcal{S}^{n}$, one needs to introduce the quantities of the first fundamental form (or, the so-called projection map) and the second fundamental form (or, the so-called extrinsic curvature), which are defined as follows:
\begin{equation}
P_{\mu\nu}:=g_{\mu\nu}
+n_{\mu}n_{\nu}\,,
\label{ProjectionMap}
\end{equation}
and
\begin{equation}
K_{\mu\nu}:=\frac{1}{2}\mathcal{L}_{n}P_{\mu\nu}\,,
\label{ExtrinsicCurvature}
\end{equation}
respectively, where $\mathcal{L}_{n}$ is the Lie derivative operator with respect to the normal vector $n^{\mu}$. Using these quantities, the Gauss equation holds:
\begin{equation}
{^{(n)}\lc{R}}{}^{L}{}_{IJK}=P^{L}{}_{\sigma}P^{\mu}{}_{I}P^{\nu}{}_{J}P_{K}{}^{\rho}\lc{R}{}^{\sigma}_{\ \mu\nu\rho}
-2K^{L}{}_{[I}K_{J]K}\,,
\label{GaussEq}
\end{equation}
where $K_{IJ}$ is the so-called extrinsic curvature and it is 
given as follows:
\begin{equation}
K_{IJ}=-\frac{
1}{2N}\left(2\lc{D}_{[I}N_{J]}-\dot{h}_{IJ}\right)\,,
\label{ExtrinsicCurvatureOnLeaf}
\end{equation}
where $\lc{D}_{I}$ is the covariant derivative of the Christoffel symbols on a leaf $\Sigma_{t}$. Contracting indices in the Gauss equation, we get the following identity
\begin{equation}
{^{(n)}\lc{R}}=\lc{R}
+\left(2\lc{R}_{\mu\nu}n^{\mu}n^{\nu}+K^{IJ}K_{IJ}-K^{2}\right)\,.
\label{ContractedGaussEq}
\end{equation}
Therefore, applying these equations the Einstein-Hilbert action in Eq~$(\ref{EH action})$ is decomposed as follows:
\begin{equation}
S_{\rm EH}=\int_{\mathcal{I}}dt\int_{\Sigma_{t}}dx^{n}N\sqrt{h}\left({^{(n)}\lc{R}}-K^{2}+K^{IJ}K_{IJ}\right)-2\int_{\mathcal{M}}d^{n+1}x\sqrt{-g}\lc\nabla_{\alpha}\left(n^{\mu}\lc\nabla_{\mu}n^{\alpha}-n^{\alpha}\lc\nabla_{\mu}n^{\mu}\right)\,.
\label{EH-actionWithBTInADM}
\end{equation}
As considered in Sec.~\ref{sec:03:02}, the boundary terms can be neglected.~\footnote{
Based on Sec.~\ref{sec:03:02}, the first term in the boundary terms can be vanished by setting properly spatial boundary conditions, and the second term can be canceled out by adding the Gibbon-York-Hawking counter-term: $-\int_{\Sigma_{t}}dx^{n}\sqrt{h}(2 
K)$ $(t\in \mathcal{I})$
.}
Therefore, the ADM-foliated Einstein-Hilbert action becomes as follows:
\begin{equation}
S_{\rm EH}=\int_{\mathcal{I}}dt\int_{\Sigma_{t}}dx^{n}N\sqrt{h}\left({^{(n)}\lc{R}}-K^{2}+K^{IJ}K_{IJ}\right):=\int_{\mathcal{I}}dt\int_{\Sigma_{t}}dx^{n}\mathcal{L}_{\rm EH}\,.
\label{EH-actionInADM}
\end{equation}
Remark that density variables and also PB-algebras are defined on a leaf $\Sigma_{t}$ $(t\in \mathcal{I})$.~\footnote{
These ingredients can be generically defined on a hypersurface such that it is homotopic to $\Sigma_{t}$ with the common spatial boundary: $\partial\Sigma_{t}$.
}

Let us perform the Dirac-Bergmann analysis (See Appendix~\ref{sec:03:01}). The canonical momentum variables are calculated as follows:
\begin{equation}
\pi_{0}:=\frac{\delta\mathcal{L}_{\rm EH}}{\delta\dot{N}}=0\,,\ \ \ \pi_{I}:=\frac{\delta\mathcal{L}_{\rm EH}}{\delta\dot{N}^{I}}=0\,,\ \ \ 
\pi_{IJ}:=\frac{\delta\mathcal{L}_{\rm EH}}{\delta\dot{h}^{IJ}}=\sqrt{h}\left(Kh_{IJ}-K_{IJ}\right)\,.
\label{CMsOfGR}
\end{equation}
Therefore, the Hessian matrix has its rank of $n(n+1)/2$. This indicates that there are four primary constraint densities as follows:
\begin{equation}
\phi^{(1)}_{0}:=\pi_{0}:\approx0\,,\quad \phi^{(1)}_{I}:=\pi_{I}:\approx0\,.
\label{PriConstsInGR}
\end{equation}
These constraint densities restrict the whole phase space to the subspace $\mathfrak{C}^{(1)}$. The total Hamiltonian density is given as follows:
\begin{equation}
\mathcal{H}_{T}=N\mathcal{C}^{\rm (GR)}_{0}+N^{I}\mathcal{C}^{\rm (GR)}_{I}+\lambda^{\mu}\phi^{(1)}_{\mu}:=\mathcal{H}_{0}+\lambda^{\mu}\phi^{(1)}_{\mu}\,,
\label{THInGR}
\end{equation}
where $\lambda^{\mu}$ are Lagrange multipliers, and $\mathcal{C}^{\rm (GR)}_{\mu}$ are defined as follows:
\begin{equation}
\mathcal{C}^{\rm (GR)}_{0}:=-\sqrt{h}{^{(n)}\lc{R}}+\frac{1}{\sqrt{h}}\left(\pi^{IJ}\pi_{IJ}-\frac{1}{n-1}\pi^{2}\right)\,,\quad 
\mathcal{C}^{\rm (GR)}_{I}:=-2\lc{D}{}^{J}\pi_{IJ}\,,
\label{C0CiInGR}
\end{equation}
where we neglected the spatial boundary term $2\pi^{IJ}N_{J}$ on ${\partial\Sigma_{t}}$ $(t\in \mathcal{I})$. $\pi$ is the trace of $\pi_{IJ}$. The fundamental PB-algebras are given as follows:
\begin{equation}
\{N(x),\pi_{0}(y)\}=\delta^{(3)}(\vec{x}-\vec{y})\,,\ \ \ \{N^{I}(x),\pi_{J}(y)\}=\delta^{I}_{J}\delta^{(3)}(\vec{x}-\vec{y})\,\,,\ \ \ \{h^{IJ}(x),\pi_{KL}(y)\}=2\delta^{(I}_{K}\delta^{J)}_{L}\delta^{(3)}(\vec{x}-\vec{y})\,.
\label{FPB-algeras}
\end{equation}
The consistency conditions for the primary constraint densities $\phi^{(1)}_{\mu}$, {\it i\,.e\,.,} $\dot{\phi}^{(1)}_{\mu}=\{\phi^{(1)}_{\mu},\mathcal{H}_{T}\}:\approx0$, give four secondary constraint densities as follows:
\begin{equation}
\phi^{(2)}_{0}:=\mathcal{C}^{\rm (GR)}_{0}:\approx0\,\,,\ \ \ \phi^{(2)}_{I}:=\mathcal{C}^{\rm (GR)}_{I}:\approx0\,.
\label{SecConstsInGR}
\end{equation}
These constraint densities further restrict $\mathfrak{C}^{(1)}$ to the new subspace $\mathfrak{C}^{(2)}$. Notice that $\phi^{(1)}_{\mu}$ and $\phi^{(2)}_{\mu}$ do not contain the lapse function and the shift vector. Therefore, all these constraint densities  are commutative with respect to the Poisson bracket. In such case, to investigate the consistency conditions for the secondary constraint densities  $\phi^{(2)}_{\mu}$, it is convenient to consider the smeared variables of $\phi^{(2)}_{\mu}$:
\begin{equation}
C_{S}(N):=\int_{\Sigma_{t}}dx^{3}
N\mathcal{C}^{\rm (GR)}_{0}\,,\quad C_{V}(\vec{N}):=\int_{\Sigma_{t}}dx^{3}
N^{I}\mathcal{C}^{\rm (GR)}_{I}\,,
\label{SmearedVariables}
\end{equation}
where $\vec{N}:=N^{I}\partial_{I}$. Then we can show that 
\begin{equation}
\{F(h^{IJ},\pi_{IJ}),C_{V}(\vec{N})\}=\mathcal{L}_{\vec{N}}F(h^{IJ},\pi_{IJ})
\label{FormulaOfCVInGR}
\end{equation}
for arbitrary function $F(h^{IJ},\pi_{IJ})$, therefore, the following algebras hold:
\begin{equation}
\{C_{V}(\vec{N}_{1}),C_{V}(\vec{N}_{2})\}=C_{V}(\mathcal{L}_{\vec{N}_{1}}\vec{N}_{2})\,,\ \ \ \{C_{V}(\vec{N}),C_{S}(N)\}=C_{S}(\mathcal{L}_{\vec{N}}N)\,.
\label{SmearedPBofCVCS}
\end{equation}
Neglecting spatial boundary terms, the following algebra holds:
\begin{equation}
\{C_{S}(N_{1}),C_{S}(N_{2})\}=C_{V}(N_{1}\partial^{I}N_{2}-N_{2}\partial^{I}N_{1})\,.
\label{SmearedPBofCS}
\end{equation}
Using these smeared PB-algebras, it can show that the consistency conditions for the secondary constraint densities are satisfied without any additional conditions and there are no tertiary constraint densities. Therefore, the analysis stops here and there are eight first-class constraints. This indicates that GR has
\begin{equation}
{\rm pDoF}=\left[2\times\frac{(n+1)(n+2)}{2}-2\times0-2\times\{(n+1)+(n+1)\}\right]\times\frac{1}{2}=\frac{1}{2}(n+1)(n-2)\,.
\end{equation} 
Notice that the multipliers remain arbitrary. This implies the existence of 
\begin{equation}
{\rm gDoF}=n+1\,.
\end{equation}
In fact, GR has the diffeomorphism invariance. In particular, in the $(3+1)$-dimensional spacetime, GR has two pDoF and four gDoF. 

If we refrain from utilizing smeared variables and instead express our results in terms of density variables, we can arrive at the following algebraic expressions:
\begin{equation}
    \{\phi^{(2)}_{I}(x),\phi^{(2)}_{J}(y)\}=\left(\phi^{(2)}_{J}(x)\partial^{(x)}_{I}-\phi^{(2)}_{I}(y)\partial^{(y)}_{J}\right)\delta^{(n)}(\vec{x}-\vec{y})\,,\quad \{\phi^{(2)}_{I}(x),\phi^{(2)}_{0}(y)\}=\phi^{(2)}_{0}(x)\partial^{(x)}_{I}\delta^{(n)}(\vec{x}-\vec{y})
\label{DensityPBofCVCS}
\end{equation}
instead of Eq~(\ref{SmearedPBofCVCS}), and
\begin{equation}
    \{\phi^{(2)}_{0}(x),\phi^{(2)}_{0}(y)\}=\left(h^{IJ}(x)\phi^{(2)}_{J}(x)+h^{IJ}(y)\phi^{(2)}_{J}(y)\right)\partial^{(x)}_{I}\delta^{(n)}(\vec{x}-\vec{y})
\label{DensityPBofCS}
\end{equation}
instead of Eq~(\ref{SmearedPBofCS}). Here, the problematic term ``$\sqrt{h}A(x)\partial^{(x)}_{I}\delta^{(n)}(\vec{x}-\vec{y})$" appears both in Eqs~(\ref{DensityPBofCVCS}) and~(\ref{DensityPBofCS}). Fortunately, the coefficients of these PBs are composed only of the secondary constraints. Therefore, without applying the prescription given in Sec.~\ref{sec:03:02}, these PBs are weakly equal to zero on $\mathfrak{C}^{(2)}$. The consistency conditions for $\phi^{(2)}_{\mu}\approx0$ has a similar property, although it is not trivial and extra calculations are mandatory differing from the case using the smeared variables. That is, it is just convenient to use smeared variables for calculating the consistency conditions when existing only first-class constraints. If there are second-class constraints, the smeared variables just make all calculations complicated due to the absence of a closed algebra on the entire phase space. There is no reason to use the smeared variables if there are second-class constraints. For instance, the authors in~\cite{Blagojevic2020} do not use the smeared variables but density variables in their analysis. In fact, it is hard to get insight into whether or not second-class constraints exist in complicated theories such as $f(T)$-gravity in advance.

GR can be extended into a non-linear manner as follows~\cite{Buchdahl1970}:
\begin{equation}
S_{f(\lc{R})}:=\int_{\mathcal{M}}d^{n+1}x\sqrt{-g}f(\lc{R})\,,
\label{fEH-action}
\end{equation}
where $f'(\lc{R})=df(\lc{R})/d\lc{R}$ and $f''\neq0$. Introducing an auxiliary field $\varphi$, Eq $(\ref{fEH-action})$ is decomposed as follows:
\begin{equation}
S_{f(\lc{R})}=\int_{\mathcal{M}}d^{n+1}x\sqrt{-g}\left[f'(\varphi)\lc{R}+f(\varphi)-\varphi f'(\varphi)\right]\,.
\label{fEH-actionInAuxiliaryVariable}
\end{equation}
Using the Gauss equation Eq~(\ref{ContractedGaussEq}) to decompose $\lc{R}$, this action becomes as follows:
\begin{equation}
S_{f(\lc{R})}=\int_{\mathcal{I}}dt\int_{\Sigma_{t}}d^{n}xN\sqrt{h}\left[f'\left({^{(n)}\lc{R}}-K^{2}+K^{IJ}K_{IJ}-\varphi\right)+f\right]
-2\int_{\mathcal{M}}d^{n+1}x\sqrt{-g}\left[f'\lc\nabla_{\mu}\left(n^{\nu}\lc\nabla_{\nu}n^{\mu}-n^{\mu}\lc\nabla_{\nu}n^{\nu}\right)\right]\,.
\label{fEH-actionWithBTInADM}
\end{equation}
Comparing to Eq~(\ref{EH-actionWithBTInADM}), the boundary terms cannot be neglected due to the existence of the non-linearity of $f'$. Integrating by parts and then neglecting the boundary terms,~\footnote{Based on Sec. \ref{sec:03:02}, the boundary term that cannot be vanished by spatial boundary conditions is canceled out by introducing the Gibbons-York-Hawking term~\cite{Alhamawi2019}: $-\int_{\Sigma_{t}}dx^{n}\sqrt{h}(2 
f'K)$ $(t\in \mathcal{I})$
.} Eq~(\ref{fEH-actionWithBTInADM}) becomes as follows~\cite{Liang2017}:
\begin{equation}
\begin{split}
S_{f(\lc{R})}=\int_{\mathcal{I}}dt\int_{\Sigma_{t}}d^{n}xN\sqrt{h}&\left[f'\left({^{(n)}\lc{R}}-K^{2}+K^{IJ}K_{IJ}-\varphi\right)+f\right] \\
&+\int_{\mathcal{I}}dt\int_{\Sigma_{t}}d^{n}xN\sqrt{h}\left[\frac{2K}{N}\left(N^{I}\lc{D}_{I}f'-f''\dot{\varphi}\right)+2\lc{D}_{I}f'{\lc{D}}{}^{I}{\rm ln}N\right]\,.
\end{split}
\label{fEH-actionWithoutBTInADM}
\end{equation}
The canonical momentum variables are calculated as follows~\cite{Liang2017}:
\begin{equation}
\pi_{0}:=0\,,\ \ \ \pi_{I}:=0\,,\ \ \ \pi_{IJ}:=\sqrt{h}\left[f'\left(Kh_{IJ}-K_{IJ}\right)-\frac{h_{IJ}}{N}\left(N^{K}\lc{D}_{K}f'-f''\dot{\varphi}\right)\right]\,,\ \ \ \pi_{\varphi}:=-2K\sqrt{h}f''\,.
\label{CMsOffGR}
\end{equation}
In the case of $f'=constant$, as expected, Eq~$(\ref{CMsOffGR})$ becomes Eq~(\ref{CMsOfGR}). Therefore, the Hessian matrix has its rank of $n(n+1)/2+1$. The primary constraint densities are given as follows:
\begin{equation}
\phi^{(1)}_{0}:=\pi_{0}:\approx0\,\,,\ \ \ \phi^{(1)}_{I}:=\pi_{I}:\approx0\,,
\label{PriConstsInfGR}
\end{equation}
and these constraint densities identify the subspace $\mathfrak{C}^{(1)}$. The total Hamiltonian density is calculated as follows:
\begin{equation}
\mathcal{H}_{T}=N\mathcal{C}^{f(\lc{R})}_{0}+N^{I}\mathcal{C}^{f(\lc{R})}_{I}+\lambda^{\mu}\phi^{(1)}_{\mu}:=\mathcal{H}_{0}+\lambda^{\mu}\phi^{(1)}_{\mu}\,
\label{THInfGR}
\end{equation}
where $\lambda^{\mu}$ are Lagrange multipliers, and $\mathcal{C}^{f(\lc{R})}_{\mu}$ are defined as follows:
\begin{equation}
\begin{split}
\mathcal{C}^{f(\lc{R})}_{0}:=&-\sqrt{h}\left[f+f'\left({^{(n)}\lc{R}}-\varphi\right)\right]+\frac{1}{\sqrt{h}f'}\left(\pi^{IJ}\pi_{IJ}-\frac{1}{n-1}\pi^{2}\right)+2\sqrt{h}\lc{D}_{I}\lc{D}{}^{I}f'-\frac{1}{n\sqrt{h}f''}\pi\pi_{\varphi}+\frac{n-1}{n\sqrt{h}f'}\left(\frac{f'}{f''}\right)^{2}\pi_{\varphi}^{2},\\
\mathcal{C}^{f(\lc{R})}_{I}:=&\pi_{\varphi}\lc{D}_{I}
\varphi
-2\lc{D}{}^{J}
\pi_{IJ}
\,.
\end{split}
\label{C0CiInfGR}
\end{equation}
In the case of $f'=constant$, Eq~(\ref{C0CiInfGR}) coincides with Eq~(\ref{C0CiInGR}). The consistency conditions for the primary constraint densities $\phi^{(1)}_{\mu}$ are gives four secondary constraint densities as follows:
\begin{equation}
\phi^{(2)}_{0}:=\mathcal{C}^{f(\lc{R})}_{0}:\approx0\,,\ \ \ \phi^{(2)}_{I}:=\mathcal{C}^{f(\lc{R})}_{I}:\approx0\,,
\label{SecConstsInfGR}
\end{equation}
under the same fundamental PB-algebras Eq~(\ref{FPB-algeras}) and
\begin{equation}
\{\varphi(x),\pi_{\varphi}(y)\}=\delta^{(3)}(\vec{x}-\vec{y})\,.
\label{ExtendedFPB-algeras}
\end{equation}
These constraint densities restrict $\mathfrak{C}^{(1)}$ to the new subspace $\mathfrak{C}^{(2)}$. We can show that the consistency conditions for these secondary constraint densities are automatically satisfied in the same manner as the GR case. That is, the smeared algebras which are given in Eqs~(\ref{SmearedPBofCVCS}) and~(\ref{SmearedPBofCS}) hold just replacing $\mathcal{C}^{\rm (GR)}_{\mu}$ by $\mathcal{C}^{f(\lc{R})}_{\mu}$ in Eq $(\ref{SmearedVariables})$.~\footnote{
Eq $(\ref{FormulaOfCVInGR})$ is generalised into $\{F(h^{IJ},\pi_{IJ},\varphi,\pi_{\varphi}),C_{V}(\vec{N})\}=\mathcal{L}_{\vec{N}}F(h^{IJ},\pi_{IJ},\varphi,\pi_{\varphi})$. The consideration in terms of density variables is viable also for the $f(\lc{R})$-gravity case.
} Therefore, $f(\lc{R})$-gravity has 
\begin{equation}
{\rm pDoF}=\frac{1}{2}(n+1)(n-2)+1\,,\quad \textrm{and}\quad {\rm gDoF}=n+1\,.
\label{}
\end{equation} 
In particular, in the $(3+1)$-dimensional spacetime, $f(\lc{R})$-gravity has three pDoF and four gDoF. 

When comparing $f(\lc{R})$-gravity with GR there is a notable property; both the theories have the common PB-algebras. Since the PB-algebras construct the generator of gauge transformation by combining as $G:=\zeta^{\mu}_{s}\phi^{(s)}_{\mu}$ $(s\in\{1,2\}; \mu\in\{1,2,\cdots,n+1\})$ for arbitrary functions $\zeta^{\mu}_{s}$ that are defined in the whole phase space, the property indicates that $f(\lc{R})$-gravity departs only of the pDoF from GR~\cite{Kimura1990,Sugano1990}. That is, $f(\lc{R})$-gravity is a natural extension of GR as unchanging the gauge symmetry. This result is consistent with the fact that $f(\lc{R})$-gravity is equivalent to the scalar-tensor theories~\cite{John1972,Teyssandier1983}.


\section{\label{sec:04}Hamiltonian analysis of Coincident GR}
In this section, as the final preparation for the main purpose of the current paper, we review the ADM-foliation of STEGR in the coincident gauge denoted by CGR and its Dirac-Bergmann analysis while remarking on the consideration given in Sec.~\ref{sec:03:02}. 

\subsection{\label{sec:04:01}ADM-foliation of Coincident GR}
The action of CGR is already derived in Sec.~\ref{sec:02:04} as Eq~(\ref{CGR-action}). The action can be rewritten as follows:
\begin{equation}
S_{CGR}=\int_{\mathcal{M}}d^{n+1}x\sqrt{-g}\frac{1}{4}M^{\alpha\beta\sigma\rho\mu\nu}Q_{\alpha\beta\sigma}Q_{\rho\mu\nu}\,,
\label{ArrangedCGR-action}
\end{equation}
where $M^{\alpha\beta\sigma\rho\mu\nu}$ is set as follows:
\begin{equation}
M^{\alpha\beta\sigma\rho\mu\nu}:=g^{\alpha\rho}g^{\beta\sigma}g^{\mu\nu}-g^{\alpha\rho}g^{\beta\mu}g^{\sigma\nu}+2g^{\alpha\nu}g^{\beta\mu}g^{\sigma\rho}-2g^{\alpha\beta}g^{\mu\nu}g^{\sigma\rho}\,.
\label{M-metric}
\end{equation}
Remark that $Q_{\alpha\beta\gamma}=\nabla_{\alpha}g_{\beta\gamma}$ is now in the coincident gauge: $Q_{\alpha\beta\gamma}=\partial_{\alpha}g_{\beta\gamma}$. Applying the ADM-foliated metric Eq~(\ref{ADM-metric}) and, after performing very long but straightforward algebraic calculations, the above action can be rewritten as follows:~\footnote{We used Cadabra to derive this result~\cite{Kasper2007}.}
\cite{Fabio2020}
\begin{equation}
S_{CGR}=\int_{\mathcal{I}}dt\int_{\Sigma_{t}}d^{n}x\sqrt{h}\left[N\left({^{(n)}Q}+K^{IJ}K_{IJ}-K^{2}\right)+\mathcal{B}_{1}+\mathcal{B}_{2}+\mathcal{B}_{3}\right]\,,
\label{ArrangedCGR-actionWithBT}
\end{equation}
where ${^{(n)}Q}$, $\mathcal{B}_{1}$, $\mathcal{B}_{2}$, and $\mathcal{B}_{3}$ are set as follows:
\begin{equation}
{^{(n)}Q}:=\frac{1}{4}\left[-h^{AD}h^{BE}h^{CF}+2h^{AE}h^{BD}h^{CF}+h^{AE}h^{BD}h^{CF}+h^{AD}h^{BC}h^{EF}+2h^{AB}h^{CD}h^{EF}\right]Q_{ABC}Q_{DEF}\,,
\label{n-dimQ}
\end{equation}
\begin{equation}
\begin{split}
\mathcal{B}_{1}=&h^{IJ}h^{KL}(Q_{JKL}-Q_{KJL})\partial_{I}N\,,\\
\mathcal{B}_{2}=&K\partial_{I}N^{I}+\dot{N}\frac{\partial_{I}N^{I}}{N^{2}}-\frac{(\partial_{I}N^{I})(N^{J}\partial_{J}N)}{N^{2}}\,,\\
\mathcal{B}_{3}=&\frac{(N^{I}\partial_{J}N)(\partial_{I}N^{J})}{N^{2}}-\frac{\partial_{I}N^{J}}{2N}\left(2\partial_{J}N^{I}+N^{I}h^{MN}Q_{JMN}\right)+\dot{N}^{K}\frac{1}{2N^{2}}(Nh^{IJ}Q_{KIJ}-2\partial_{K}N)\,.
\end{split}
\label{}
\end{equation}
The boundary terms $\mathcal{B}_{1}$, $\mathcal{B}_{2}$, and $\mathcal{B}_{3}$ are calculated, neglecting spatial boundary terms based on the consideration in Sec.~\ref{sec:03:02}, respectively, as follows: 
\begin{equation}
\mathcal{B}_{1}=-N\sqrt{h}\lc{D}_{I}({^{(n)}Q}^{I}-{^{(n)}\tilde{Q}}^{I})\,\,,\ \ \ \mathcal{B}_{2}=\partial_{\mu}N^{J}\partial_{J}(\sqrt{h}n^{\mu})\,\,,\ \ \ \mathcal{B}_{3}=-\partial_{I}N^{I}\partial_{\mu}(\sqrt{h}n^{\mu})\,.
\label{}
\end{equation}
Therefore, the action becomes as follows:
\begin{equation}
S_{CGR}=\int_{\mathcal{I}}dt\int_{\Sigma_{t}}d^{n}x\left[N\sqrt{h}\left\{{^{(n)}Q}+K^{IJ}K_{IJ}-K^{2}-\lc{D}_{I}({^{(n)}Q}^{I}-{^{(n)}\tilde{Q}}^{I})\right\}-\partial_{I}N^{I}\partial_{\mu}(\sqrt{h}n^{\mu})+\partial_{\mu}N^{J}\partial_{J}(\sqrt{h}n^{\mu})\right]\,.
\label{ArrangedCGR-actionWithoutBT}
\end{equation}
This is none other than the ADM-foliation of CGR~\cite{Fabio2020}. Notice that the derivation of Eq $(\ref{ArrangedCGR-actionWithoutBT})$ neglected only spatial boundary terms. Integrating the second term in Eq $(\ref{ArrangedCGR-actionWithoutBT})$ by parts: $-\partial_{I}N^{I}\partial_{\mu}(\sqrt{h}n^{\mu})=-\partial_{I}[N^{I}\partial_{\mu}(\sqrt{h}n^{\mu})]+\partial_{\mu}[N^{I}\partial_{I}(\sqrt{h}n^{\mu})]-(\partial_{\mu}N^{I})(\partial_{I}\sqrt{h}n^{\mu})$, the third term is canceled out with the third term in Eq~(\ref{ArrangedCGR-actionWithoutBT}). The remnant terms are only boundary terms; these terms can be neglected based on the consideration in Sec.~\ref{sec:03:02}. Therefore, we get the final result:
\begin{equation}
S_{CGR}=\int_{\mathcal{I}}dt\int_{\Sigma_{t}}d^{n}xN\sqrt{h}\left[{^{(n)}Q}+K^{IJ}K_{IJ}-K^{2}-\lc{D}_{I}({^{(n)}Q}^{I}-{^{(n)}\tilde{Q}}^{I})\right]\,.
\label{CGR-actionWithoutBT}
\end{equation}
Note that the above derivation does not need the Gauss equation unlike the GR and $f(\lc{R})$-gravity cases. We just manipulated complicated algebraic calculations. Remark, finally, that this neglection of the boundary terms needs more careful consideration when extending the theory in a non-linear manner like the $f(\lc{R})$-gravity case.

\subsection{\label{sec:04:02}Hamiltonian analysis of Coincident GR
}
For the action Eq~(\ref{CGR-actionWithoutBT}), we perform the Dirac-Bergmann analysis. The canonical momentum variables are calculated as follows:
\begin{equation}
\pi_{0}=0\,,\ \ \ \pi_{I}=0\,,\ \ \ \pi_{IJ}=\sqrt{h}(Kh_{IJ}-K_{IJ})\,,
\label{CMsOfCGR}
\end{equation}
and therefore, the primary constraint densities are given as follows:
\begin{equation}
\phi^{(1)}_{0}:=\pi_{0}:\approx0\,,\ \ \ \phi^{(1)}_{I}:=\pi_{I}:\approx0\,.
\label{PriConstsInCGR}
\end{equation}
The rank of the Hessian matrix is $n(n+1)/2$. These constraint densities restrict the whole phase space to the subspace $\mathfrak{C}^{(1)}$. The total Hamiltonian density is derived as follows:
\begin{equation}
\mathcal{H}_{T}:=N\mathcal{C}^{\rm (CGR)}_{0}+N^{I}\mathcal{C}^{\rm (CGR)}_{I}+\lambda^{\mu}\phi^{(1)}_{\mu}\,,
\label{THInCGR}
\end{equation}
where $\mathcal{C}^{\rm (CGR)}_{0}$ and $\mathcal{C}^{\rm (CGR)}_{I}$ are set as follows:
\begin{equation}
\mathcal{C}^{\rm (CGR)}_{0}:=-\sqrt{h}\left[{^{(n)}Q}-\lc{D}_{I}({^{(n)}Q}{}^{I}-{^{(n)}\tilde{Q}}{}^{I})\right]+\frac{1}{\sqrt{h}}\left(\pi^{IJ}\pi_{IJ}-\frac{1}{n-1}\pi^{2}\right)\,,\quad 
\mathcal{C}^{\rm (CGR)}_{I}:=-2\lc{D}{}^{J}\pi_{IJ}\,.
\label{C0CiInCGR}
\end{equation}
The fundamental PB-algebras are the same as those of GR: Eq~(\ref{FPB-algeras}). Therefore, the consistency conditions for the primary constraint densities Eq~(\ref{PriConstsInCGR}) 
lead to four secondary constraint densities:
\begin{equation}
\phi^{(2)}_{0}:=\mathcal{C}^{\rm (CGR)}_{0}:\approx0\,,\ \ \ \phi^{(2)}_{I}:=\mathcal{C}^{\rm (CGR)}_{I}:\approx0\,,
\label{SecConstsInCGR}
\end{equation}
and these constraint densities restrict $\mathfrak{C}^{(1)}$ to the new subspace $\mathfrak{C}^{(2)}$. We can show that the smeared PB-algebras of Eq~(\ref{SecConstsInCGR}) satisfy the common algebras that are those of GR given in Eqs~(\ref{SmearedPBofCVCS}) and~(\ref{SmearedPBofCS}) after tedious calculations along with neglecting properly spatial boundary terms~\cite{Fabio2020}, and it indicates that CGR has the same gauge symmetry as GR and $f(\lc{R})$-gravity. That is, the analysis stops here. Therefore, CGR has
\begin{equation}
{\rm pDoF}=\frac{1}{2}(n+1)(n-2)\,,\quad \textrm{and}\quad {\rm gDoF}=n+1\,.
\label{}
\end{equation} 
In particular, $(3+1)$-dimensional spacetime, CGR has two pDoF and four gDoF. That is, CGR is completely equivalent to GR from both viewpoints of dynamics and gauge symmetry, as expected. 

\section{\label{sec:05}Hamiltonian analysis of Coincident $f(Q)$-gravity}
In this section, after reviewing the controversy in $f(T)$-gravity and providing our perspective of it and the role of the prescription proposed in Sec.~\ref{sec:03:02}, we perform the analysis of coincident $f(Q)$-gravity. It reveals that, as a generic case, the theory has five primary, three secondary, and two tertiary constraint densities, and all these constraint densities are classified into second-class density; the pDoF and gDoF of the theory are six and zero, respectively. 

\subsection{\label{sec:05:00}A lesson from the Dirac-Bergmann analysis of $f(T)$-gravity}
As mentioned in Sec.~\ref{sec:01}, there was a controversy on the pDoF of $f(T)$-gravity due to the existence of second-class constraint densities. It implies that some of multipliers are determined, but, here, a problematic situation occurs. That is, a set of Partial Differential Equations (PDEs) of Lagrange multipliers, which has been not expected in the Dirac-Bergmann analysis in point particle systems and at least in GR, fGR, and CGR, appears. In addition, the existence is a feature for violating the diffeomorphism and/or local Lorentz symmetry. It implies that the system has several sectors of solutions and each sector generically has different pDoF. In this case, the Dirac-Bergmann analysis gives rise to different results depending on assumptions. In order to see these issues, focusing in particular on the determination of the multipliers, let us briefly review the case of $f(T)$-gravity in four-dimensional spacetime. Since the coincident $f(Q)$-gravity has also second-class constraint densities, this quick survey gives an insight into the use of the prescription given in Sec.~\ref{sec:03:02}. 

As mentioned in Sec.~\ref{sec:01}, there are three works; (i) Li {\it et al.}~\cite{Li2011}: pDoF is five; (ii) Ferraro and Guzmán~\cite{Ferraro2018}: pDoF is three; (iii) Blagojevic and Nester~\cite{Blagojevic2020}: pDoF is five as a generic case. In these works, in order to count out the pDoF, the Dirac-Bergmann analysis is commonly applied. However, the methods to derive constraint densities and determine Lagrange multipliers are different. In (i) Li {\it et al.}~\cite{Li2011} and (ii) Ferraro and Guzmán~\cite{Ferraro2018}, on one hand, the rank of the Dirac matrix is investigated to find constraint densities and determine Lagrange multipliers. On the other hand, in (iii) Blagojevic and Nester~\cite{Blagojevic2020}, Castellani's algorithm~\cite{Castellani:1981us} is applied to find first-class constraint densities. For deriving second-class constraint densities and determining Lagrange multipliers, an original method is applied, as briefly reviewed later. The important point here is that these approaches lead to the common first-class constraint densities including its PB-algebras under the imposition of the constraint densities, and the PB-algebras are nothing but that of general relativity, which are already given in Eqs~(\ref{DensityPBofCVCS}) and~(\ref{DensityPBofCS}). That is, $f(T)$-gravity is a diffeomorphism invariant theory, as expected. 

Next, let us see a different point among these works. This is, the emergence of the second-class constraint densities and the determination of the multipliers. This difference leads to the two different results in the pDoF of $f(T)$-gravity as a generic case. The first survey is (i) Li {\it et al.}~\cite{Li2011}: pDoF is five. In their work, the primary second-class constraint densities are derived as follows:
\begin{equation}
    \Gamma^{ab}:\approx0\,,\quad \pi:\approx0\,,
\label{2nd-class constraints in Li etal}
\end{equation}
where $a\,,b\in\{0\,,1\,,2\,,3\}$, $\Gamma^{ab}$ and $\pi$ are canonical momentum variables with respect to the vielbein and auxiliary field, respectively. The auxiliary field is necessary for decomposing the Lagrangian of $f(T)$-gravity in the same manner as the case of $f(\lc{R})$-gravity. The PBs among $\Gamma^{ab}$ and $\pi$ do not vanish under the imposition of the constraint densities. The concrete forms of the algebras are given in~\cite{Li2011}. Then the consistency conditions for Eq~(\ref{2nd-class constraints in Li etal}) are derived as follows:
\begin{equation}
    M\Lambda:\approx0\,,
\label{Consistency conditions for 2nd-class constraints in Li etal}
\end{equation}
where $\Lambda$ is a column vector with eight components and the Dirac matrix $M$ is given as follows:
\begin{equation}
M=\left(
\begin{array}{cccccccc}
 0 & y_1 & y_2 & y_3 & y_4 & y_5 & y_6 & x_0\\
 -y_1 & 0 & 0 & 0 & A_{11} & A_{12} & A_{13} & x_1 \\
 -y_2 & 0 & 0 & 0 & A_{21} & A_{22} & A_{23} & x_2 \\
 -y_3 & 0 & 0 & 0 & A_{31} & A_{32} & A_{33} & x_3 \\
 -y_4 & -A_{11} & -A_{21} & -A_{32} & 0 & B_{12} & B_{13} & x_4\\
 -y_5 & -A_{12} & -A_{22} & -A_{32} & -B_{12} & 0 & B_{23} & x_5\\
 -y_6 & -A_{13} & -A_{23} & -A_{33} & -B_{13} & -B_{23}  & 0  & x_6\\
 -x_0 & -x_1 & -x_2 & -x_3 & -x_4 & -x_5 & -x_6 & 0
\end{array}
\right)\,,
\label{Matrix M in Li etal}
\end{equation}
where $A_{ij}$ and $B_{ij}$ are proportional to $\partial_{k}\varphi$ $(i\,,j\,,k\in\{1\,,2\,,3\})$, where $\varphi$ is an auxiliary field. The concrete forms of $\Lambda$, $A_{ij}$, $B_{ij}$, and other variables, $x_{0}\,,x_{1}\,,\cdots\,,x_{6}\,,;\,,y_{0}\,,y_{1}\,,\cdots\,,y_{6}$, are not important in this survey. The explicit formulae are given in~\cite{Li2011}. Remark, here, that, since the first-class constraint densities are commutative with the second-class constraint densities by its definition under the imposition of all the constraint densities, it is enough to consider the consistency conditions for the second-class constraint densities~\cite{Li2011}. In order to exist a nontrivial solution of Eq~(\ref{Consistency conditions for 2nd-class constraints in Li etal}), the determinant of $M$ has to vanish, and it gives rise to a new constraint density as follows:
\begin{equation}
    \pi_{1}=\sqrt{{\rm det}\,M}:\approx0\,.
\label{Secondary constraint in Lorentz sector in Li etal}
\end{equation}
$\pi_{1}$ is proportional to $(\partial_{i}\varphi)^{3}$~\cite{Li2011}. Then the matrix $M$ has its rank of six. This means that six out of seven multipliers are determined. The consistency condition for $\pi_{1}$ determines the remaining multiplier. Then, we can show that the extended matrix of $M$ taking into account the PBs of $\pi_{1}$ with the other constraint densities has its rank of eight by performing the elementary transformation of matrices~\cite{Li2011}. Therefore, eight first-class and eight second-class constraints exist, and then the pDoF is $(34-8\times2-8)/2=5$. 

An important point here is that the PB of $\pi$ and $\pi_{1}$ becomes the following form:
\begin{equation}
    \{\pi(x)\,,\pi_{1}(y)\}\approx\,\propto(\partial_{i}\varphi)^{3}\delta^{(3)}(\vec{x}-\vec{y})\,+\propto(\partial_{i}\varphi)^{2}\partial_{i}\delta^{(3)}(\vec{x}-\vec{y})\,,
\label{PB of pi and pi1 in Li et al}
\end{equation}
schematically, where we denoted ``$\,\propto{\cdots}\,$'' as a term which is proportional to ``$\,\cdots\,$''. Notice that the problematic term ``$\sqrt{h}A(x)\partial^{(x)}_{I}\delta^{(n)}(\vec{x}-\vec{y})$'' appears. Even if the prescription given in Sec.~\ref{sec:03:02} is applied and the problematic term is neglected under the imposition of the spatial boundary condition: $N_{i}(t\,,{\rm spatial\ boundary}):=0$, $\pi_{1}$ still determines the remaining one multiplier with respect to the second constraint in Eq~(\ref{2nd-class constraints in Li etal}) by virtue of the first term in Eq~(\ref{PB of pi and pi1 in Li et al}); the procedure also stops here, and this case is also the same pDoF as five. We will discuss this point later. 

The second survey is (ii) Ferraro and Guzmán~\cite{Ferraro2018}: pDoF is three. This number contradicts to the result of Li {\it et al.}~\cite{Li2011}, but we will discuss the reason for this point later. In their work~\cite{Li2011}, the primary second-class constraint densities are derived as follows:
\begin{equation}
    G^{(1)}_{ab}:\approx0\,,\quad G^{(1)}_{\pi}:\approx0\,
\label{2nd-class constraints in Ferraro etal}
\end{equation}
with respect to the vielbein and auxiliary field, respectively, where $a\,,b\in\{0\,,1\,,2\,,3\}$. These constraint densities are the same as those of Li {\it et al.}~\cite{Li2011} excepting the notations. The Dirac matrix is also equivalent to that of Li {\it et al.}~\cite{Li2011} excepting the notations and the inclusion of the first-class constraint densities (but it is not mandatory since the first-class constraint densities are commutative with the other constraint densities). Differing from the method to count the rank of the Dirac matrix, {\it i\,.e\,.,} the fundamental transformation of matrices, which is used in Li {\it et al.}~\cite{Li2011}, the authors utilized the method of using null eigenvectors to find new secondary constraint densities. However, this difference is not crucial since these methods are mathematically equivalent. The difference point between these works is the composition of the primary constraint densities. That is, the authors reconstructed the primary constraint densities $G^{(1)}_{ab}$ excepting for $G^{(1)}_{01}$ as follows:
\begin{equation}
\begin{split}
    \tilde{G}^{(1)}_{02}=&F_{01}G^{(1)}_{02}-F_{02}G^{(1)}_{01}\,,\\
    \tilde{G}^{(1)}_{03}=&F_{02}G^{(1)}_{03}-F_{03}G^{(1)}_{01}\,,\\
    &\vdots\\
    \tilde{G}^{(1)}_{23}=&F_{01}G^{(1)}_{23}-F_{23}G^{(1)}_{01}\,,
\end{split}
\label{Reconstructed G^1_ab in Ferraro etal}
\end{equation}
where $F_{ab}$ $(a\,,b\in\{0\,,1\,,2\,,3\})$ are composed of the vielbein and those spatial derivatives. The explicit forms are given in~\cite{Ferraro2018}. As shown in~\cite{Ferraro2018}, the PBs between $G^{(1)}_{\pi}$ and $\tilde{G}^{(1)}_{ab}$ vanish excepting for $\tilde{G}^{(1)}_{01}:=G^{(1)}_{01}$. That is, before reconstructing $G^{(1)}_{ab}$, the PBs between $G^{(1)}_{ab}$ and $G^{(1)}_{\pi}$ are not commutative, but now so are in those of $\tilde{G}^{(1)}_{ab}$ and $G^{(1)}_{\pi}$.\footnote{
This result can be understood by considering the following instance. Let us consider a set of second-class constraints: $p_{1}\approx0$, $p_{2}+q^{1}\approx0$, and $q^{1}\approx0$. If the third constraint is applied to the second one then the latter one becomes $p_{2}\approx0$; we obtained a new set of constraints: $p_{1}\approx0$, $p_{2}\approx0$, and $q^{1}\approx0$. In this set, the second constraint is classified into first-class while remaining the first and third ones second-class constraints. This is a special feature of second-class constraints. In the case of first-class constraints, since these constraints form a Lie algebra of the gauge symmetry of a given system, there is no such feature. 
}
In addition, the authors performed further reconstruction as follows:
\begin{equation}
    \tilde{G}^{(2)}_{0}=F_{01}G^{(2)}_{0}-F_{\varphi}G^{(1)}_{01}\,,
\label{Reconstructed G^2_0 in Ferraro etal}
\end{equation}
where $G^{(2)}_{0}$ is the secondary first-class constraint density with respect to one of the Hamiltonian constraint density $G^{(1)}_{0}$ in the sector of diffeomorphism symmetry. $F_{\varphi}$ is composed of the torsion, auxiliary field, and vielbein. The explicit forms are given in~\cite{Ferraro2018}. Then the PB between $\tilde{G}^{(2)}_{0}$ and $G^{(1)}_{\pi}$ becomes commutative under the imposition of $G^{(1)}_{0}\approx0$. Therefore, the authors concluded that only $G^{(1)}_{\pi}$ and $\tilde{G}^{(1)}_{01}$ are classified into second-class constraint densities; pDoF is $(34-(8+5)\times2-2)/2=3$. 

Here, let us discuss the relation between (i) Li {\it et al.}~\cite{Li2011} and (ii) Ferraro and Guzmán~\cite{Ferraro2018}. The difference of the pDoF in these works (and (iii) Blagojevic and Nester~\cite{Blagojevic2020}) is noting but the controversy in the analysis of $f(T)$-gravity. The point to resolve this situation is that the proportionality of $\partial_{k}\varphi$ in $A_{ij}$ and $B_{ij}$. That is, if we assume that the configuration $\varphi$ is independent from the space coordinates, {\it i\,.e\,.,} $\varphi=\varphi(t)$, then the terms $A_{ij}$ and $B_{ij}$ vanish. In this case, the rank of $M$ becomes two, and only two multipliers are determined. Then the determinant of $M$ given in Eq~(\ref{Secondary constraint in Lorentz sector in Li etal}) is automatically satisfied; $\pi_{1}$ is no longer a constraint density. Therefore, under the restriction of the configuration $\varphi$ to $\varphi=\varphi(t)$, there are thirteen first-class constraint densities and two secondary constraint densities, and this situation precisely corresponds to the result of (ii) Ferraro and Guzmán~\cite{Ferraro2018}. That is, (ii) Ferraro and Guzmán~\cite{Ferraro2018} is a sector of (i) Li {\it et al.}~\cite{Li2011} with the specific configuration of $\varphi=\varphi(t)$.

The final survey is the work (iii) Blagojevic and Nester~\cite{Blagojevic2020}: pDoF is five as a generic case. This work provides the most detailed results: there are five sectors, {\it i\,.e\,.,} (s1) pDoF is five as a generic case; (s2) pDoF is N/A (a detailed investigation is necessary); (s3) pDoF is four as a special case; (s4) pDoF is two as a generic case; (s5) pDoF is two as a special case. All these sectors have different composition of constraint densities. Of course, (i) Li {\it et al.}~\cite{Li2011} belongs to the sector (s1). We will discuss the case of (ii) Ferraro and Guzmán~\cite{Ferraro2018} later. In their work~\cite{Blagojevic2020}, the primary second-class constraint densities are derived as follows:
\begin{equation}
    \pi_{\varphi}:\approx0\,,\quad C_{ij}:\approx0\,,
\label{2nd-class constraints in Blagojevic etal}
\end{equation}
schematically, where $i\,,j\in\{0\,,1\,,2\,,3\}$. These constraint densities are the same as (i) Li {\it et al.}~\cite{Li2011} and (ii) Ferraro and Guzmán~\cite{Ferraro2018} excepting the notations. The consistency conditions for these constraint densities, $\chi=\dot{\pi}_{\varphi}:\approx0$ and $\chi_{ij}=\dot{C}_{ij}:\approx0$, determine some of the multipliers. According to the authors, the latter conditions are split into two parts with respect to the time and space direction of the ADM-foliation, and then the analysis is classified into two main sectors whether or not $\varphi_{\bar{i}}:=\partial_{\bar{i}}\varphi=\partial_{i}\varphi-n_{i}n^{j}\partial_{j}\varphi\neq0$ holds. In the main sector of $\varphi_{\bar{i}}\neq0$, the ADM-foliated consistency conditions lead to a new secondary constraint density denoted as $\chi:\approx0$. Taking into account the consistency condition for $\chi$, the authors decompose the seven multipliers such that the determination of a specific multiplier, denoted as $u$, also determines all the other multipliers. The authors derive an equation as follows:
\begin{equation}
    u(x)D(x\,,x')=G(x')\,,
\label{Eq of u in Blagojevic etal}
\end{equation}
where $D(x\,,x')$ and $G(x')$ are composed only of the phase space variables and the spatial derivatives of these variables. In particular, $D(x\,,x')$ contains the spatial derivative of $\delta$-function. This implies that the problematic term ``$\sqrt{h}A(x)\partial^{(x)}_{I}\delta^{(n)}(\vec{x}-\vec{y})$'' appears. Therefore, Eq~(\ref{Eq of u in Blagojevic etal}) leads to the following equation:
\begin{equation}
    A^{\gamma}\partial_{\gamma}u+\alpha u=G\,,
\label{PDE of u in Blagojevic etal}
\end{equation}
schematically, where $\gamma\in\{1\,,2\,,3\}$. Based on this result, the authors classify this case into the following three possible sectors; (s1) {\it If the PDE~(\ref{PDE of u in Blagojevic etal}) is solvable} then pDoF is $(34-8\times2-8)/2=5$; (s2) If both $A^{\gamma}$ and $\alpha$ vanish then there gives rise to further constraint densities (but any specific result are not derived due to the difficulty of the computations of PBs); (s3) If the consistency condition for $\chi$ is automatically satisfied then pDoF is $(34-(8+2)\times2-6)=4$. In particular, in Sector (s1), notice that even if the prescription given in Sec.~\ref{sec:03:02} is applied and the problematic terms are neglected under the imposition of the spatial boundary condition: $N_{i}(t\,,{\rm spatial\ boundary}):=0$, Eq~(\ref{PDE of u in Blagojevic etal}) determines the multiplier $u$ by an algebraic equation. In other words, the prescription guarantees the solvability of Eq~(\ref{PDE of u in Blagojevic etal}). We will discuss this point later. In the main sector of $\varphi_{\bar{i}}=0$, the authors lead to the following two sectors: (s4) If $\varphi=constant$ then the system in turn results in a system of TEGR with a cosmological constant term, that is, ${\rm pDoF}=2$; (s5) Seven new constraint densities appear and all of them are classified into second-class constraint densities, that is, pDoF is $(34-8\times2-14)/2=2$.  

Here, let us discuss the relation between (ii) Ferraro and Guzmán~\cite{Ferraro2018} and (iii) Blagojevic and Nester~\cite{Blagojevic2020}. Since (ii) Ferraro and Guzmán~\cite{Ferraro2018} is equivalent to (i) Li {\it et al.}~\cite{Li2011} with the specific configuration $\varphi=\varphi(t)$, (ii) Ferraro and Guzmán~\cite{Ferraro2018} is classified into the main sector of $\varphi_{\bar{i}}=0$ in (iii) Blagojevic and Nester~\cite{Blagojevic2020}. However, both Sectors (s4) and (s5) are not the case of (ii) Ferraro and Guzmán~\cite{Ferraro2018} due to the difference in the constraint structures. In order to resolve this situation, let us reconsider (ii) Ferraro and Guzmán~\cite{Ferraro2018}. The point is that $\tilde{G}^{(1)}_{01}=G^{(1)}_{01}$ contains a term being proportional to $\varphi$~\cite{Ferraro2018}. This term prevent to make the PB between $G^{(1)}_{\pi}$ and $\tilde{G}^{(1)}_{01}$ commutative. That is, if the auxiliary field $\phi$ becomes a constant then $G^{(1)}_{\pi}$ and $\tilde{G}^{(1)}_{01}$ turn into first-class constraint densities, and then pDoF becomes $(34-(8+7)\times2)/2=2$. This is nothing but Sector (s4) in (iii) Blagojevic and Nester~\cite{Blagojevic2020}. Remark that in this sector the local Lorentz invariance is restored; all the constraint densities are now first-class constraint densities. Therefore, (ii) Ferraro and Guzmán~\cite{Ferraro2018} is noting but a generic case of Sector (s4) in (iii) Blagojevic and Nester~\cite{Blagojevic2020}. In fact, the author assume an additional condition $\dot{\varphi}\approx0$; this means of course that $\varphi$ is a constant. This implies that one out of three degrees of freedom in (ii) Ferraro and Guzmán~\cite{Ferraro2018} should be some sort of ghost degrees of freedom. (It propagates/dynamical but is unphysical.) 

Finally, let us consider the role of the prescription (See Sec.~\ref{sec:03:02}) in the analysis of $f(T)$-gravity. The point to grasp the truth of the role is that $f(T)$-gravity is a diffeomorphism invariant theory. That is, this means that the theory does not depend on a coordinate choice, or equivalently, an ADM-foliation. An ADM-foliation is determined from the normal vector, denoted as $n^{\mu}$, of leafs. (See Sec.~\ref{sec:03:03}.) Explicitly, it was expressed as $n^{\mu}=-N^{-1}(-1\,,N^{I})$, where $N$ and $N^{I}$ are a lapse function and a shift vector. The inverse, $n_{\mu}$, is $n_{\mu}=-N(1\,,N_{I})$ with the satisfaction of $N^{I}N_{I}=0$. Therefore, the theory is invariant in any choice of a lapse function and a shift vector, and the application of the prescription, which demands the spatial boundary condition of vanishing shift vector at least on the spatial boundary, does not change the theory anything. That is, for a diffeomorphism invariant theory, without any loss of generality, the prescription can apply to the Dirac-Bergmann analysis. This is the reason why the ignorance of the problematic term ``$\sqrt{h}A(x)\partial^{(x)}_{I}\delta^{(n)}(\vec{x}-\vec{y})$'' by imposing the spatial boundary condition $N_{I}(t\,,{\rm spatial\ boundary}):=0$ did not change the result of the analysis in Eqs~(\ref{PB of pi and pi1 in Li et al}) and~(\ref{Eq of u in Blagojevic etal}) respectively. Whereas, in the case of the coincident $f(Q)$-gravity, however, the situation gets changed; Hu {\it et al.}~\cite{Katsuragawa2022} unveiled that Eqs~(\ref{DensityPBofCVCS}) and~(\ref{DensityPBofCS}), which are the algebra of diffeomorphism invariance of a gravity theory, are at least partly violated. This means that the prescription cannot apply in a generic manner, differing from the case of $f(T)$-gravity. Nevertheless, the prescription has an advantage by virtue of the following reason: the circumventing of the PDEs of Lagrange multipliers. In fact, in the work of Hu {\it et al.}~\cite{Katsuragawa2022}, this point was overlooked and indicated by D’Ambrosio {\it et al.}~\cite{Fabio2023} with the statement that the pDoF should be up to seven. In our perspective, since the coincident $f(Q)$-gravity is not diffeomorphism invariant, therefore, it has several sectors just being analogous to the violation of the Lorentz invariance in $f(T)$-gravity, and each sector generically has a different pDoF one another. In other words, Hu {\it et al.}~\cite{Katsuragawa2022} unveiled the pDoF of a possible generic sector in the coincident $f(Q)$-gravity, that is, pDoF is eight, although the issue of the solvability of the PDEs of the multiplier remains. In our work, in order to circumvent this problem, we perform the Dirac-Bergmann analysis under the imposition of the prescription, meaning that we apply the spatial boundary condition $N_{I}(t\,,{\rm spatial\ boundary}):=0$ not only to the variational principle but also to the PB-algebras. That is, we will investigate the other main sector, which is different from the main sector investigated by Hu {\it et al.}~\cite{Katsuragawa2022}, and then we will obtain six pDoF as a generic case. Now, let us move on to the main thesis of the current paper.

\subsection{\label{sec:05:01}ADM-foliation of Coincident $f(Q)$-gravity}
We performed the Dirac-Bergmann analysis of GR, CGR, and $f(\lc{R})$-gravity in a $(n+1)$-dimensional spacetime. As we will see, however, the existence of the second-class constraint densities makes it difficult to understand whether the consistency conditions determine the multipliers or derive new constraint densities since the size of the Dirac matrix becomes bigger. Therefore, in this section, for simplicity, we perform the analysis for the coincident $f(Q)$-gravity in a $(3+1)$-dimensional spacetime and then estimate the general case of the dimension of $(n+1)$. A general proof would be completed by applying the mathematical induction.

In the same manner as the case of $f(\lc{R})$-gravity, the CGR can be extended non-linearly into as follows:
\begin{equation}
S_{f(Q)}=\int_{\mathcal{M}}d^{4}x\sqrt{-g}f(Q)\,,
\end{equation}
where $f$ is an arbitrary function of the nonmetricity scalar. By introducing an auxiliary variable we obtain:
\begin{equation}
S_{f(Q)}=\int_{\mathcal{M}}d^{4}x\sqrt{-g}\left[f'Q+f-\varphi f'\right]\,,
\label{fCGR1}
\end{equation}
where $f$ is an arbitrary function of the auxiliary variable $\varphi$ and $f'':=d^{2}f/d\varphi^{2}\neq0$. From Sec.~\ref{sec:04:01}, using the ADM-foliation of $Q$ given in Eq~(\ref{ArrangedCGR-actionWithoutBT}), Eq~(\ref{fCGR1}) can be decomposed as follows:
\begin{equation}
\begin{split}
S_{f(Q)}=\int_{\mathcal{I}}dt\int_{\Sigma_{t}}d^{3}x
&
\left[N\sqrt{h}f'\left\{{^{(3)}Q} +K^{IJ}K_{IJ}-K^{2}-\lc{D}_{I}\left({^{(3)}Q}^{I}-{^{(3)}Q}^{I}\right)\right\}+N\sqrt{h}\left(f-\varphi f'\right)\right.\\
&
\left.+f'\left\{\partial_{\mu}N^{I}\partial_{I}\left(\sqrt{h}n^{\mu}\right)-\partial_{I}N^{I}\partial_{\mu}\left(\sqrt{h}n^{\mu}\right)\right\}+\sqrt{h}f'\lc{D}_{I}\left\{N\left({^{(3)}Q}^{I}-{^{(3)}Q}^{I}\right)\right\}\right]\,.
\end{split}
\label{}
\end{equation}
Remark that this foliation takes all boundary terms into account. Integrating by parts and neglecting the spatial boundary terms by imposing appropriate spatial boundary conditions on $\partial\Sigma_{t}$, we get
\begin{equation}
\begin{split}
S_{f(Q)}=\int_{\mathcal{I}}dt\int_{\Sigma_{t}}d^{3}x
&
\left[N\sqrt{h}\left\{f'\left({^{(3)}Q} +K^{IJ}K_{IJ}-K^{2}\right)-\lc{D}_{I}\left\{f'\left({^{(3)}Q}^{I}-{^{(3)}Q}^{I}\right)\right\}\right\}+f-\varphi f'\right\}\\
&
\left.+f'\left\{\partial_{\mu}N^{I}\partial_{I}\left(\sqrt{h}n^{\mu}\right)-\partial_{I}N^{I}\partial_{\mu}\left(\sqrt{h}n^{\mu}\right)\right\}\right]\,.
\end{split}
\label{}
\end{equation}
Further, integrating by parts the first term of and the second term of the boundary term with respect to the spatial derivative and the spacetime derivative, respectively, and neglecting each the boundary term on $\partial\Sigma_{t}$ and $\partial\mathcal{M}$, respectively, we obtain the following formula:
\begin{equation}
\begin{split}
S_{f(Q)}=\int_{\mathcal{I}}dt\int_{\Sigma_{t}}d^{3}x
&
\left[N\sqrt{h}\left\{f'\left({^{(3)}Q} +K^{IJ}K_{IJ}-K^{2}\right)-\lc{D}_{I}\left\{f'\left({^{(3)}Q}^{I}-{^{(3)}Q}^{I}\right)\right\}\right\}+f-\varphi f'\right\}\\
&
\left.+\frac{\sqrt{h}}{N}\left(N^{I}\partial_{J}N^{J}-N^{J}\partial_{J}N^{I}\right)\partial_{I}f'-\frac{\sqrt{h}}{N}\left(\partial_{I}f'\right)\dot{N}^{I}+\frac{\sqrt{h}}{N}f''\left(\partial_{I}N^{I}\right)\dot{\varphi}\right]\,.
\end{split}
\label{ADM-foliated action of f(Q)}
\end{equation}
That is, the non-linearity of $f$ changes the constraint structure of CGR. This action was first derived in~\cite{Katsuragawa2022} by a different method that resembles the GR case. The canonical momentum variables are computed as follows:
\begin{equation}
\begin{split}
\pi_{0}:=&\frac{\delta S_{f(Q)}(x)}{\delta \dot{N}(y)}=0,\\
\pi_{I}:=&\frac{\delta S_{f(Q)}(x)}{\delta \dot{N}^{I}(y)}=-\frac{\sqrt{h}}{N}f''\partial_{I}\varphi\delta^{(3)}(\vec{x}-\vec{y}),\\
\pi_{IJ}:=&\frac{\delta S_{f(Q)}(x)}{\delta \dot{h}^{IJ}(y)}=\sqrt{h}f'\left(K_{IJ}-Kh_{IJ}\right)\delta^{(3)}(\vec{x}-\vec{y}),\\
\pi_{\varphi}:=&\frac{\delta S_{f(Q)}(x)}{\delta \dot{\varphi}(y)}=\frac{\sqrt{h}}{N}f''\partial_{I}N^{I}\delta^{(3)}(\vec{x}-\vec{y})\,.
\end{split}
\label{}
\end{equation}
The canonical momentum variables with respect to the shift vectors depart from the ordinary CGR case; it depends on the lapse function, the 3-metric $h_{IJ}$, and the non-linearity part by $f''$. The canonical momentum with respect to the auxiliary variable $\varphi$ generates a constraint density, which is different from the case of $f(\lc{R})$-gravity. That is, in coincident $f(Q)$-gravity, the auxiliary variable $\varphi$ does not have any physical feature unlike $f(\lc{R})$-gravity. 

\subsection{\label{sec:05:02:01}Primary constraint densities and total Hamiltonian density}
The Hessian matrix of the system has the size of $11\times 11$ - components only being non-vanishing components with respect to the canonical momenta $\pi_{IJ}$. All other components of the matrix vanish. Therefore, the rank of the Hessian matrix is six, and it implies that the system has five primary constraint densities given as follows:
\begin{equation}
\begin{split}
\phi^{(1)}_{0}:=&\pi_{0}:\approx0\,,\\
\phi^{(1)}_{I}:=&\pi_{I}+\frac{\sqrt{h}}{N}f''\partial_{I}\varphi:\approx0\,,\\
\phi^{(1)}_{\varphi}:=&\pi_{\varphi}-\frac{\sqrt{h}}{N}f''\partial_{I}N^{I}:\approx0\,.
\end{split}
\label{primary constraints}
\end{equation}
These constraint densities restrict the whole phase space to the subspace $\mathfrak{C}^{(1)}$. The PB-algebras among these primary constraint densities are computed as follows:
\begin{equation}
\begin{split}
\{\phi^{(1)}_{0}(x),\phi^{(1)}_{I}(y)\}=
&
\frac{1}{N^{2}}\sqrt{h}f''\partial_{I}\varphi\delta^{(3)}(\vec{x}-\vec{y}):=A_{I}\delta^{(3)}(\vec{x}-\vec{y})\,,\\
\{\phi^{(1)}_{0}(x),\phi^{(1)}_{\varphi}(y)\}=
&
-\frac{1}{N^{2}}\sqrt{h}f''\partial_{I}N^{I}\delta^{(3)}(\vec{x}-\vec{y}):=B\delta^{(3)}(\vec{x}-\vec{y})\,,\\
\{\phi^{(1)}_{I}(x),\phi^{(1)}_{\varphi}(y)\}=
&
\frac{1}{N}\sqrt{h}f'''\partial_{I}\varphi\delta^{(3)}(\vec{x}-\vec{y}):=C_{I}\delta^{(3)}(\vec{x}-\vec{y})\,,
\end{split}
\label{PBsOfPriConsts}
\end{equation}
where we neglected all spatial boundary terms respecting the discussion in Sec.~\ref{sec:03:02}. Therefore, these five primary constraint densities are classified into second-class constraint density. 

Explicitly, the problematic term occurs in the third PB given in Eq~(\ref{PBsOfPriConsts}) as follows:
\begin{equation}
 \{\phi^{(1)}_{I}(x),\phi^{(1)}_{\varphi}(y)\}:=C_{I}\delta^{(3)}(\vec{x}-\vec{y})+\left[\left(\frac{\sqrt{h(x)}}{N(x)}f''(x)\right)\partial^{(x)}_{I}+\left(\frac{\sqrt{h(y)}}{N(y)}f''(y)\right)\partial^{(y)}_{I}\right]\delta^{(3)}(\mathbf{x}-\mathbf{y})\,.
\label{PBsOfPriConstsOfIAndVarphi}
\end{equation}
Taking into account that $\phi^{(1)}_{I}(x)$ and $\phi^{(1)}_{\varphi}(y)$ are density variables, in order to reveal the meaning of the above equations in mathematically correct manner, we have to integrate it out with respect to $d^{3}x$ and $d^{3}y$ on a leaf $\Sigma_{t}$. Then, the first term in Eq~(\ref{PBsOfPriConstsOfIAndVarphi}) turns into the smeared one, and the second terms becomes generically as follows:
\begin{equation}
\begin{split}
    &\int_{\Sigma_{t}}\int_{\Sigma_{t}}\left[\left.\left(\frac{\sqrt{h}}{N}f''\right)\right|_{x}\partial^{(x)}_{I}\delta^{(3)}(\mathbf{x}-\mathbf{y})\right]dx^{3}dy^{3}\\
    &=\int_{\Sigma_{t}}\int_{\Sigma_{t}}\left[\partial^{(x)}_{I}\left(\frac{\sqrt{h}}{N}f''\delta^{(3)}(\mathbf{x}-\mathbf{y})\right)-\partial^{(x)}_{I}\left(\frac{\sqrt{h}}{N}f''\right)\delta^{(3)}(\mathbf{x}-\mathbf{y})\right]dx^{3}dy^{3}\\
    &=-\left.N_{I}\sqrt{h}f''\left(1-\mathfrak{D}(\mathbf{x}\rightleftarrows\mathbf{y})\right)\right|_{\partial\Sigma_{t}}\,,
\end{split}
\label{IntegratedPBsOfPriConstsOfIAndVarphi}
\end{equation}
where we assumed $n_{\mu}=(-N\,,NN_{I})$ with $N^{I}N_{I}=0$ and $\mathfrak{D}(\mathbf{x}\rightleftarrows\mathbf{y})$ denotes the Lebesgue integration of the second term. If the naive replacement of ``$x$'' in the spatial derivative by ``$y$'' is possible then $\mathfrak{D}(\mathbf{x}\rightleftarrows\mathbf{y})=1$, and Eq~(\ref{IntegratedPBsOfPriConstsOfIAndVarphi}) vanishes without any additional condition. However, if it is not the case, then we fix the shift vectors $N_{I}=0$ as the spatial boundary condition then the problematic term is removed from the analysis. In fact, the action integral Eq~(\ref{ADM-foliated action of f(Q)}) contains only the first-order time derivative of $N_{I}$, and it indicates that $N_{I}$ cannot be any physical variable; it affects only the constraint structure of the system. Otherwise, it would also be possible to impose $f''=0$ on $\partial\Sigma_{t}$. (If $\varphi^{2}\in f''$ then we modify the term $\varphi^{2}$ as $\alpha\varphi^{2}$ with $\alpha\rightarrow0$ on $\partial\Sigma_{t}$). Hereinafter, we use this prescription to set well-defined PBs, which is formulated in detail in Sec.~\ref{sec:03:02}. Where, ``{\it well-defined}'' means that, integrated over the PBs with respect to all space variables, the PBs are composed only of the terms that are proportional to $\delta$-function.

The Legendre transformation of the coincident $f(Q)$-gravity is calculated as follows:
\begin{equation}
\mathcal{H}_{0}:=N\mathcal{C}^{f(Q)}_{0}+N^{I}\mathcal{C}^{f(Q)}_{I}\,,
\label{H0}
\end{equation}
where $\mathcal{C}^{f(Q)}_{0}$ and $\mathcal{C}^{f(Q)}_{I}$ are defined as follows:
\begin{equation}
\begin{split}
\mathcal{C}^{f(Q)}_{0}:=
&
-\sqrt{h}\left[f'{^{(3)}Q}-\lc{D}_{I}\left\{f'\left({^{(3)}Q}^{I}-{^{(3)}Q}^{I}\right)\right\}+f-\varphi f'-\frac{1}{hf'}\left(\pi^{II}\pi_{IJ}-\frac{1}{2}\pi^{2}\right)\right],\\
\mathcal{C}^{f(Q)}_{I}:=
&
-2\lc{D}^{J}\pi_{IJ}-\frac{\sqrt{h}}{N}f''\left(\partial_{J}N^{J}\partial_{I}\varphi-\partial_{I}N^{J}\partial_{J}\varphi\right)\,.
\end{split}
\label{C and Ci}
\end{equation}
Therefore, the total Hamiltonian density of the system is introduced as follows:
\begin{equation}
\mathcal{H}_{T}:=\mathcal{H}_{0}+\lambda_{0}\phi^{(1)}_{0}+\sum^{3}_{I=1}\lambda_{I}\phi^{(1)}_{I}+\lambda_{\varphi}\phi^{(1)}_{\varphi}\,.
\label{HT}
\end{equation}
The PB-algebras among the primary constraint densities and the density $\mathcal{H}_{0}$ are given in Appendix~\ref{App01}. 

\subsection{\label{sec:05:02:02}Consistency conditions for primary constraint densities and the emergence of secondary constraint densities}
The consistency conditions for the primary constraint densities $\phi^{(1)}_{\alpha}$ are given as follows:
\begin{equation}
\dot{\phi}^{(1)}_{\alpha}:=\{\phi^{(1)}_{\alpha},\mathcal{H}_{T}\}\approx\{\phi^{(1)}_{\alpha},\mathcal{H}_{0}\}+\lambda_{\beta}\{\phi^{(1)}_{\alpha},\phi^{(1)}_{\beta}\}:\approx0\,,
\label{ConsistencyOfPrimaryConsts}
\end{equation}
where $\alpha,\beta$ run in the range of $\alpha,\beta\in\{0,1,2,3,\varphi\}$. The appearance of the same indices in the formula means applying Einstein's summation convention. Since all these primary constraint densities are classified into second-class constraint density, it is necessary to investigate the rank of the Dirac matrix $D^{(1)}_{\alpha\beta}\delta^{(3)}(\vec{x}-\vec{y}):=\{\phi^{(1)}_{\alpha},\phi^{(1)}_{\beta}\}$:
\begin{equation}
D^{(1)}:=\begin{bmatrix}
0 & A_{1} & A_{2} & A_{3} & B \\
-A_{1} & 0 & 0 & 0 & C_{1} \\
-A_{2} & 0 & 0 & 0 & C_{2} \\
-A_{3} & 0 & 0 & 0 & C_{3} \\
-B & -C_{1} & -C_{2} & -C_{3} & 0 
\end{bmatrix}\,,
\label{D1}
\end{equation}
where $A_{I}$, $C_{I}$, and $B$ are defined by Eq~(\ref{PBsOfPriConsts}). Applying the fundamental matrix transformations to Eq~(\ref{D1}), we get the following matrix:
\begin{equation}
D'^{(1)}:=P^{(1)}D^{(1)}Q^{(1)}=\begin{bmatrix}
0 & 0 & 0 & 0 & B \\
0 & 0 & A_{12} & A_{13} & 0 \\
0 & -A_{12} & 0 & A_{23} & 0 \\
0 & -A_{13} & -A_{23} & 0 & 0 \\
-B & 0 & 0 & 0 & 0 
\end{bmatrix}\,,
\label{D1Trans}
\end{equation}
where we set $A_{IJ}:=2A_{[I}C_{J]}$. $P^{(1)}$ and $Q^{(1)}$ are set as follows:
\begin{equation}
P^{(1)}:=\begin{bmatrix}
1 & 0 & 0 & 0 & 0 \\
-\frac{C_{1}}{B} & 1 & 0 & 0 & -\frac{A_{1}}{B} \\
-\frac{C_{2}}{B} & 0 & 1 & 0 & -\frac{A_{2}}{B} \\
-\frac{C_{3}}{B} & 0 & 0 & 1 & -\frac{A_{3}}{B} \\
0 & 0 & 0 & 0 & 1
\end{bmatrix}\,\,,\ \ \ 
Q^{(1)}:=\begin{bmatrix}
1 & -\frac{C_{1}}{B} & -\frac{C_{2}}{B} & -\frac{C_{3}}{B} & 0 \\
0 & 1 & 0 & 0 & 0 \\
0 & 0 & 1 & 0 & 0 \\
0 & 0 & 0 & 1 & 0 \\
0 & -\frac{A_{1}}{B} & -\frac{A_{2}}{B} & -\frac{A_{3}}{B} & 1 \\
\end{bmatrix}\,.
\label{P1 and Q1}
\end{equation}
The straightforward computations lead to $A_{IJ}=0$. Therefore, The Dirac matrix $D^{(1)}$ has a rank of two. This indicates that two multipliers are determined and then three secondary constrain densities appear. Using these fundamental matrices $P^{(1)}$ and $Q^{(1)}$, and the Dirac matrix $D^{(1)}$, the consistency conditions Eq~(\ref{ConsistencyOfPrimaryConsts}) becomes as follows:
\begin{equation}
P^{(1)}_{\alpha\beta}\{\phi^{(1)}_{\beta},\mathcal{H}_{0}\}+D'^{(1)}_{\alpha\beta}\lambda^{(1)}_{\beta}\delta^{(3)}(\vec{x}-\vec{y}):\approx0\,,
\label{TransConsistencyOfPrimaryConsts}
\end{equation}
where we set 
\begin{equation}
\lambda^{(1)}_{\alpha}:={Q^{(1)}}^{-1}_{\alpha\beta}\lambda_{\beta}\,.
\label{TransSecondaryMultipliers}
\end{equation}
For $\alpha=\varphi$ and $\alpha=0$, the corresponding multipliers $\lambda^{(1)}_{\varphi}$ and $\lambda^{(1)}_{0}$ are determined as follows:
\begin{equation}
\lambda^{(1)}_{\varphi}=-\frac{1}{B}\{\phi^{(1)}_{0},\mathcal{H}_{0}\}\,\,,\ \ \ \lambda^{(1)}_{0}=\frac{1}{B}\{\phi^{(1)}_{\varphi},\mathcal{H}_{0}\}\,.
\label{DeterminedTransSecondaryMultipliers}
\end{equation}
The explicit formulae of these multipliers are derived by using the formulae given in Appendix \ref{App01}. The multipliers $\lambda^{(1)}_{I}$ remain arbitrary. Converting $\lambda^{(1)}_{\alpha}$ into the original multipliers $\lambda_{\alpha}$, we get
\begin{equation}
\begin{split}
\lambda_{0}=&\frac{1}{B}\{\phi^{(1)}_{\varphi},\mathcal{H}_{0}\}-\frac{C_{I}}{B}\lambda^{(1)}_{I}\,,\\
\lambda_{\varphi}=&-\frac{1}{B}\{\phi^{(1)}_{0},\mathcal{H}_{0}\}-\frac{A_{I}}{B}\lambda^{(1)}_{I}\,,\\
\lambda_{I}=&\lambda^{(1)}_{I}\,.
\end{split}
\label{SecondaryMultipliers}
\end{equation}
The secondary constraint densities are derived in the correspondence to the undetermined multipliers $\lambda_{I}=\lambda^{(1)}_{I}$ $(I\in\{1,2,3\})$:
\begin{equation}
\phi^{(2)}_{I}:=P^{(1)}_{I\alpha}\{\phi^{(1)}_{\alpha},\mathcal{H}_{0}\}=-\frac{C_{I}}{B}\{\phi^{(1)}_{0},\mathcal{H}_{0}\}+\{\phi^{(1)}_{I},\mathcal{H}_{0}\}-\frac{A_{I}}{B}\{\phi^{(1)}_{\varphi},\mathcal{H}_{0}\}:\approx0\,.
\label{SecondaryConstraints}
\end{equation}
The explicit formulae of these secondary constraint densities can be derived by using the formulae given in Appendix~\ref{App01} and it reveals that all the secondary constraint densities are classified into second-class constraint density. These constraint densities restrict $\mathfrak{C}^{(1)}$ to the new subspace $\mathfrak{C}^{(2)}$.

Utilizing Eq~(\ref{SecondaryMultipliers}), the multipliers $\lambda_{0}$ and $\lambda_{\varphi}$ in the total Hamiltonian density Eq~(\ref{HT}) are replaced by $\lambda_{I}$:
\begin{equation}
\mathcal{H}_{T}=\mathcal{H}^{(2)}_{0}+\lambda_{I}\Phi^{(2)}_{I}\,,
\label{SecondaryHT}
\end{equation}
where $\mathcal{H}^{(2)}_{0}$ and $\Phi^{(2)}_{I}$ are set as follows:
\begin{equation}
\begin{split}
\mathcal{H}^{(2)}_{0}:=&\mathcal{H}_{0}-\frac{1}{B}\{\phi^{(1)}_{0},\mathcal{H}_{0}\}\phi^{(1)}_{\varphi}+\frac{1}{B}\{\phi^{(1)}_{\varphi},\mathcal{H}_{0}\}\phi^{(1)}_{0}\\
\Phi^{(2)}_{I}:=&\phi^{(1)}_{I}-\frac{A_{I}}{B}\phi^{(1)}_{\varphi}-\frac{C_{I}}{B}\phi^{(1)}_{0}\approx0\,.
\end{split}
\label{TransH02AndPriConsts}
\end{equation}
In the next section, we calculate the tertiary constraint densities.

\subsection{\label{sec:05:02:03}Consistency conditions for secondary constraint densities and the emergence of tertiary constraint densities}
The consistency conditions for the secondary constraint densities $\phi^{(2)}_{I}$ are given as follows:
\begin{equation}
\dot{\phi}^{(2)}_{I}=\{\phi^{(2)}_{I},\mathcal{H}^{(2)}_{0}\}+\lambda_{J}\{\phi^{(2)}_{I},\Phi^{(2)}_{J}\}:\approx0\,.
\label{ConsistencyOfSecondaryConsts}
\end{equation}
The existence of tertiary constraint densities depends on the rank of the matrix $D^{(2)}_{IJ}\delta^{(3)}(\vec{x}-\vec{y}):=\{\phi^{(2)}_{I},\Phi^{(2)}_{J}\}$. Using the formulae in Appendix~\ref{App01}, we get the following result:
\begin{equation}
D^{(2)}_{IJ}=\partial_{I}\varphi(\alpha\partial_{J}\varphi+\Delta_{J})\,,
\label{D2InIndix}
\end{equation}
where $\Delta_{I}:=\beta_{I}^{J}\partial_{J}\varphi$ and the explicit formulae of $\alpha$ and $\beta^{I}_{J}$ are given in Appendix~\ref{App02}. In matrix form, $D^{(2)}$ is expressed as follows:
\begin{equation}
D^{(2)}=\begin{bmatrix}
\varphi_{1}(\alpha\varphi_{1}+\Delta_{1}) & \varphi_{1}(\alpha\varphi_{2}+\Delta_{2}) & \varphi_{1}(\alpha\varphi_{3}+\Delta_{3}) \\
\varphi_{2}(\alpha\varphi_{1}+\Delta_{1}) & \varphi_{2}(\alpha\varphi_{2}+\Delta_{2}) & \varphi_{2}(\alpha \varphi_{3}+\Delta_{3}) \\
\varphi_{3}(\alpha\varphi_{1}+\Delta_{1}) & \varphi_{3}(\alpha\varphi_{2}+\Delta_{2}) & \varphi_{3}(\alpha\varphi_{3}+\Delta_{3})
\end{bmatrix}\,,
\label{D2InMatrix}
\end{equation}
where $\varphi_{I}:=\partial_{I}\varphi$. Applying the fundamental matrix transformations to Eq~(\ref{D2InMatrix}), we get
\begin{equation}
D'^{(2)}:=P^{(2)}D^{(2)}Q^{(2)}=\begin{bmatrix}
\varphi_{1}(\alpha \varphi_{1}+\Delta_{1}) & 0 & 0\\
0 & 0 & 0 \\
0 & 0 & 0
\end{bmatrix}\,,
\label{D2Trans}
\end{equation}
where $P^{(2)}$ and $Q^{(2)}$ are set as follows:
\begin{equation}
P^{(2)}:=\begin{bmatrix}
1 & 0 & 0 \\
-\varphi_{2} & \varphi_{1} & 0\\
-\varphi_{3} & 0 & \varphi_{1}
\end{bmatrix}\,\,,\ \ \ 
Q^{(2)}:=\begin{bmatrix}
1 & -\frac{\alpha\varphi_{2}+\Delta_{2}}{\alpha\varphi_{1}+\Delta_{1}} & -\frac{\alpha\varphi_{3}+\Delta_{3}}{\alpha\varphi_{1}+\Delta_{1}} \\
0 & 1 & 0\\
0 & 0 & 1
\end{bmatrix}\,.
\label{P2 and Q2}
\end{equation}
Therefore, the matrix $D^{(2)}$ has its rank of one. This indicates that one multiplier is determined and then two tertiary constraint densities appear. The consistency conditions in Eq~(\ref{ConsistencyOfSecondaryConsts}) can be rewritten as follows:
\begin{equation}
P^{(2)}_{IJ}\{\phi^{(2)}_{J},\mathcal{H}^{(2)}_{0}\}+D'^{(2)}_{IJ}\lambda^{(2)}_{J}\delta^{(3)}(\vec{x}-\vec{y}):\approx0\,,
\label{TransConsistencyOfSecondaryConsts}
\end{equation}
where $\lambda^{(2)}_{I}$ is set as follows:
\begin{equation}
\lambda^{(2)}_{I}={Q^{(2)}}^{-1}_{IJ}\lambda_{J}\,.
\label{TransTertiaryMultipliers}
\end{equation}
Therefore, the multiplier to the $x$-component, $\lambda^{(2)}_{1}$, is determined as follows:
\begin{equation}
\lambda^{(2)}_{1}=-\frac{1}{
\varphi_{1}(\alpha\varphi_{1}+\Delta_{1})}P^{(2)}_{1I}\{\phi^{(2)}_{I},\mathcal{H}^{(2)}_{0}\}=-\frac{1}{
\varphi_{1}(\alpha\varphi_{1}+\Delta_{1})}\{\phi^{(2)}_{1},\mathcal{H}^{(2)}_{0}\}\,,
\label{DeterminedTransTertiaryMultiplier}
\end{equation}
where we used $P^{(2)}$ in Eq~(\ref{P2 and Q2}). The explicit formula of Eq $(\ref{DeterminedTransTertiaryMultiplier})$ can be derived by using Eq $(\ref{SecondaryConstraints})$, the first formula in Eq~(\ref{TransH02AndPriConsts}), and the formulae given in Appendix~\ref{App01}. The multipliers $\lambda^{(2)}_{I'}$ $(I'\in\{2,3\})$ remain arbitrary. Converting $\lambda^{(2)}_{I}$ into the original multipliers $\lambda_{I}$, we get
\begin{equation}
\begin{split}
\lambda_{1}=&-\frac{1}{
\varphi_{1}(\alpha\varphi_{1}+\Delta_{1})}\{\phi^{(2)}_{1},\mathcal{H}^{(2)}_{0}\}-\frac{\alpha \varphi_{2}+\Delta_{2}}{\alpha\varphi_{1}+\Delta_{1}}\lambda^{(2)}_{2}-\frac{\alpha\varphi_{3}+\Delta_{3}}{\alpha\varphi_{1}+\Delta_{1}}\lambda^{(2)}_{3}\,, \\
\lambda_{I'}=&\lambda^{(2)}_{I'}\,.
\end{split}
\label{TertiaryMultipliers}
\end{equation}
The tertiary constraint densities are derived in the correspondence to the undetermined multipliers $\lambda_{I'}=\lambda^{(2)}_{I'}$:
\begin{equation}
\phi^{(3)}_{I'}:=P^{(2)}_{I'J}\{\phi^{(2)}_{J},\mathcal{H}^{(2)}_{0}\}:\approx0\,,
\label{TertiaryConstraints}
\end{equation}
that is,
\begin{equation}
\begin{split}
\phi^{(3)}_{2}:=&-\varphi_{2}\{\phi^{(2)}_{1},\mathcal{H}^{(2)}_{0}\}+\varphi_{1}\{\phi^{(2)}_{2},\mathcal{H}^{(2)}_{0}\}:\approx0\,,\\
\phi^{(3)}_{3}:=&-\varphi_{3}\{\phi^{(2)}_{1},\mathcal{H}^{(2)}\}+\varphi_{1}\{\phi^{(2)}_{3},\mathcal{H}^{(2)}_{0}\}:\approx0\,.
\end{split}
\label{SeparatedTertiaryConstraints}
\end{equation}
The explicit formulae can be derived by using Eq~(\ref{SecondaryConstraints}), the first formula in Eq~(\ref{TransH02AndPriConsts}), and the formulae given in Appendix~\ref{App01} and it reveals that all the tertiary constraint densities are classified into second-class constraint density. These constraint densities restrict $\mathfrak{C}^{(2)}$ to the new subspace $\mathfrak{C}^{(3)}$.

Utilizing Eq~(\ref{TertiaryConstraints}), the total Hamiltonian density Eq~(\ref{SecondaryHT}) is rewritten as follows:
\begin{equation}
\mathcal{H}_{T}:=\mathcal{H}^{(3)}_{0}+\lambda_{I'}\Phi^{(3)}_{I'}\,,
\label{TertiaryHT}
\end{equation}
where $\mathcal{H}^{(3)}_{0}$ and $\Phi^{(3)}_{I'}$s are set as follows:
\begin{equation}
\begin{split}
\mathcal{H}^{(3)}_{0}:=&\mathcal{H}^{(2)}_{0}-\frac{1}{
\varphi_{1}(\alpha \varphi_{1}+\Delta_{1})}\{\phi^{(2)}_{1},\mathcal{H}^{(2)}_{0}\}\Phi^{(2)}_{1},\\
\Phi^{(3)}_{I'}:=&\Phi^{(2)}_{I'}-\frac{\alpha \varphi_{I'}+\Delta_{I'}}{\alpha\varphi_{1}+\Delta_{1}}\Phi^{(2)}_{1}\,.
\end{split}
\label{TransH03AndPriConsts}
\end{equation}
In the next section, we determine the remaining multipliers and identify the pDoF of the theory.

There is a remark: There may be a case that $\alpha \varphi_{1}+\Delta_{1}$ in Eq~(\ref{D2Trans}) vanishes, and three tertiary constraint densities appear. Since the total number of second-class constraint densities is always even, at least one more secondary constraint density exists as a quaternary or more higher order constraint density, and then the pDoF is up to $(22-5-3-3-1)/2=5$. In this case, however, there is no easy way to confirm how many constraint densities emerge due to its complexity of computations. This sort of complicated situation also appears in the analysis of $f(T)$-gravity: the sector (s2) in~\cite{Blagojevic2020}. 

\subsection{\label{sec:05:02:04}Consistency conditions for tertiary constraint densities and pDoF of coincident $f(Q)$-gravity}
The consistency conditions for the tertiary constraint densities $\phi^{(3)}_{I'}$ are given as follows:
\begin{equation}
\dot{\phi}^{(3)}_{I'}=\{\phi^{(3)}_{I'},\mathcal{H}^{(3)}_{0}\}+\lambda_{J'}\{\phi^{(3)}_{I'},\Phi^{(3)}_{J'}\}:\approx0\,.
\label{ConsistencyOfTertiaryConsts}
\end{equation}
The existence of quaternary constraints depends on the rank of the matrix $D^{(3)}_{I'J'}\delta^{(3)}(\vec{x}-\vec{y}):=\{\phi^{(3)}_{I'},\Phi^{(3)}_{J'}\}$. It is easy to confirm that the rank of $D^{(3)}$ is two (full-rank; its determinant does not vanish) since the spatial boundary terms no longer vanish without accidental cases due to that the primed indices run the range only of $2,3$, although it is very tedious to lead to the explicit formula of $D^{(3)}$ and its determinant. Therefore, the remaining multipliers are determined and then the procedure stops here excepting accidental cases. Since there are five primary, three secondary, and two tertiary constraint densities and all the constraint densities are classified into second-class constraint density, therefore, the pDoF and the gDoF of coincident $f(Q)$-gravity are
\begin{equation}
{\rm pDoF}=\frac{1}{2}\times(22-5-3-2)=6
\,,\quad \textrm{and}\quad {\rm gDoF}=0\,.
\label{}
\end{equation}

There is a remark: If the rank of $D^{(3)}$ is one then one of the remaining multipliers are determined and a quaternary constraint density appears. We already have five primary, three secondary, and two tertiary constraint densities, and all these constraint densities are classified into second-class constraint densities. It indicates that at least one more second-class constraint density has to exist as a higher order constraint density since the total number of second-class constraint densities is always an even number. Therefore, this accidental case has (b) ${\rm pDoF}\leq(22-5-3-2-1-1))/2=5$ and ${\rm gDoF=0}$. If the rank of $D^{(3)}$ is zero then at least two quaternary constraint densities appear. If these constraint densities determine all the remaining multipliers, this accidental case has (c) ${\rm pDoF}\leq(22-5-3-2-0-2)/2=5$ and ${\rm gDoF=0}$. In these cases, however, there is no easy way to confirm how many constraint densities emerge due to the complexity of computations. This sort of complicated situation also appears in the analysis of $f(T)$-gravity: the sector (s2) in~\cite{Blagojevic2020}. 

For simplicity, so far, we considered that the spacetime manifold has $3+1$ dimensions, however, it would be possible to extend this result for any spacetime dimension. This analysis would also give an implication of pDoF and gDoF. Let $\mathcal{M}$ be a $(n+1)$-dimensional spacetime manifold. To do this, we have to consider the two cases depending on $n$ is odd or even. If $n$ is an odd number then we might get $n+2$ primary, $n$ secondary, and $n-1$ tertiary constraint densities and all these constraint densities would be classified into second-class constraint densities. Therefore, when $n$ is an odd number, we have 
\begin{equation}
{\rm pDoF}=\frac{1}{2}(n^{2}+3)\,,\quad \textrm{and}\quad {\rm gDoF}=0\,,
\label{}
\end{equation}
respectively. 
If $n$ is an even number then we might get $n+2$ primary, $n$ secondary, $n-1$ tertiary, and $1$ quaternary constraint densities and all these constraint densities would be classified into second-class constraint densities. Therefore, for a $n+1$ dimensions when $n$ is an even number, we have
\begin{equation}
 {\rm pDoF}=\frac{1}{2}n^2+1\,,\quad \textrm{and}\quad {\rm gDoF}=0\,.
\end{equation}

These pDoFs are just an estimation based on the result of the analysis in the case of a $(3+1)$-dimensional spacetime. However, it would be possible to strictly prove these results by applying mathematical induction.

\section{\label{sec:06}Conclusions}
In this paper, we revise the metric-affine gauge theory of gravity by introducing the mathematical framework of the gauge approach for gravity and clarifying the positioning of the coincident GR from viewpoints of gauge fixing conditions. Then we proposed the prescription for the Dirac-Bergmann analysis to circumvent the PDEs of Lagrange multipliers. After that, we investigated the pDoF and the gDoF of GR, $f(\lc{R})$-gravity, and CGR, and we showed that the analysis did not need the prescription to count out each pDoF. The case of $f(T)$-gravity, however, the Dirac-Bergmann analysis need the prescription to guarantee the solvability of the consistency conditions on the sectors of the local Lorentz symmetry. In particular, we unveiled that the violation of a symmetry could provide several main sectors of a given theory, and we clarified the restriction of the prescription. Based on these preparations, the analysis of the coincident $f(Q)$-gravity is performed. We unveiled that, as a generic case, the coincident $f(Q)$-gravity has six propagating and zero gauge degrees of freedom; five primary, three secondary, and two tertiary constraint densities exist, and all these constraint densities are classified into second-class constraint density. The emergence of the ten second-class constraints can be interpreted as ascribing to the violation of the closed algebra composed of the eight first-class constraints of GR, which forms the generator of the diffeomorphis invariance of the theory, and the elimination of the auxiliary degrees of freedom $\varphi$ (and its canonical momentum variable $\pi_{\varphi}$). As accidental cases, there were the three sectors; (a) The pDoF is up to five: five primary, three secondary, three tertiary, and at least one higher-order constraint densities exist and all these constraint densities are classified into second-class; (b) The pDoF is up tp five: five primary, three secondary, two tertiary, one quaternary, and at least one higher-order constraint densities exist and all these constraint densities are classified into second-class; (c) The pDoF is up to five: five primary, three secondary, two tertiary, and at least two higher-order constraint densities exist and all these constraint densities are classified into second-class. These results are consistent with the upper bound of pDoF, {\it i\,.e\,.,} seven, which is claimed in~\cite{Fabio2023}. We also estimated the pDoF and gDoF of a $(n+1)$-dimensional spacetime; ${\rm pDoF}=(n^{2}+3)/2$ if $n$ is odd number and ${\rm pDoF}=n^{2}/2+1$ if $n$ is even number, and both the cases have ${\rm gDoF}=0$. However, we must emphasize that one should further investigate the pDoF of the coincident $f(Q)$-gravity, by using a new method that is proposed recently in~\cite{Kaparulin:2012px,ErrastiDiez:2020dux,ErrastiDiez:2023gme,Fabio2023}, for instance. It would be also interesting in not choosing the coincident gauge and performing the same analysis. This could clarify the assumption made here which is that the spatial boundary terms can be neglected by choosing appropriate boundary conditions implying that the Dirac-Bergmann algorithm can be used.

In the previous work in~\cite{Katsuragawa2022}, the authors derived the result that pDoF is eight without assuming any conditions in advance. This result would be the most generic case of the analysis in the coincident $f(Q)$-gravity but suffered from the solvability of the PDEs of the multipliers, which was first indicated by~\cite{Fabio2023}, it should, therefore, be regarded as a speculation. In our work, in order to circumvent this problem, we sacrificed the generality of the analysis, and then focused on a specific sector that does not suffer from the problem. Then, we obtained one generic sector and three specific sectors, of course, differing from the result in~\cite{Katsuragawa2022}. However, after once proving the solvability, the speculation would turn into a truth of the coincident $f(Q)$-gravity. Further investigation on this point is of course for a significant future work under taking into account the following remark: If the Lebesgue measure in Eq~(\ref{IntegratedPBsOfPriConstsOfIAndVarphi}) is unity, {\it i\,.e\,.,} $\mathfrak{D}(\mathbf{x}\rightleftarrows\mathbf{y})=1$, then the terms like ``$\sqrt{h}A(x)\partial^{(x)}_{I}\delta^{(3)}(\mathbf{x}-\mathbf{y})$'' in Eq~(\ref{PBsOfPriConstsOfIAndVarphi}) automatically vanish when integrating over on a leaf. In this case, another scenario is possible: the case of the absence of the prescription. Then the PB-algebras in the second term in Eq~(\ref{ConsistencyOfSecondaryConsts}) contain the spatial boundary terms, and then the matrix given in Eq~(\ref{D2InIndix}) could generically be non-singular (full-rank). This means that the tertiary constraint densities do not appear and the procedure stops here with the PDEs of the Lagrange multipliers: this is the resemble situation to the fTEGR case as already explained in Sec.~\ref{sec:05:00}. The same issue would arise also in Eqs~(\ref{ConsistencyOfPrimaryConsts}) and~(\ref{D1}). Therefore, {\it if all of these PDEs are solvable}, the pDoF becomes seven. If this scenario is realized then five primary constraints and three secondary constraint densities appear and all these constraint densities are classified into second-class constraint densities. The emergence of these eight second-class constraint densities can be interpreted as ascribing to the violation of the closed algebra composed of the eight first-class constraint densities of GR. In this case, the auxiliary variable could propagate but be unphysical degrees of freedom. This perspective would be consistent with the recent works of the cosmological perturbation~\cite{Gomes:2023hyk,Gomes:2023tur,Heisenberg:2023wgk}. As mentioned in Sec.~\ref{sec:01}, we emphasize again that it is important for cosmology to unveil the exact pDoF of our theory of each branch to know the existence of infinitely strongly coupling. If the theory is infinitely strongly coupled to each background then the standard way of linear perturbation for cosmology breaks down.

GR, CGR, and $f(\lc{R})$-gravity have only first-class constraint densities and these constraint densities satisfy the common Poisson bracket algebras (PB-algebras) but so do not for coincident $f(Q)$-gravity. This indicates that $f(\lc{R})$-gravity is a natural extension of GR; just one extra pDoF is added and the gauge symmetry does not change but coincident $f(Q)$-gravity is a departure of gravity which is described by GR; not only four extra pDoF are added but also the gauge symmetry of GR is lost. The result also indicates that only the imposition of the coincident gauge condition does not break the diffeomorphism symmetry since CGR has yet the common PB-algebras to GR until the non-linear extension is taken into account; it seems that the boundary terms have something to do with its breaking when combining the coincident gauge condition. However, this statement itself should be proven and investigated together with the well-posedness of the variational principle~\cite{Keisuke2023,Kyosuke2023} in more detail and these are also for a future work. 

The method of extension of GR is not restricted to the geometrical alternations and the non-linearization; non-localization gives a great insight into the understanding of quantum aspects of gravity \cite{Biswas2006,Capozziello2022}. In particular, it is shown that the $\lc{R}^{2}$-inflation model, which is a special class of $f(\lc{R})$-gravity, in the non-local extension is a possible candidate for a consistent effective quantum gravity theory from the viewpoint of UV-completion \cite{Stelle1978,Starobinsky1980,Koshelev2023}. It would be expected to build the resemble scenario for $f(T)$- and (coincident) $f(Q)$-gravity theories. In this regard, it would play a crucial role to distinguish these extended theories each other from viewpoints of constraint systems, as shown throughout the current paper in the local theories. For non-local $f(\lc{R})$-gravity, the Dirac-Bergmann analysis was performed in~\cite{Joshi2022}. However, the analysis for generic non-local theories has not yet been established regardless of the fact that it is expected that the analysis clarifies the differences among various non-localized theories of gravity. Constructing a general theory of the Dirac-Bergmann analysis or its alternative theory if it is necessary for non-local theories would also be for a significant future work. 

\begin{acknowledgments}
KT and SB appreciate Cadabra~\cite{Kasper2007} and Maxima~\cite{Souza2004} which were used for algebraic calculations of ADM-foliation and Dirac-Bergmann analysis. We would like to thank Jorge Gigante Valcarcel, Keisuke Izumi, Masahide Yamaguchi, Shin'ichi Hirano, and Teruaki Suyama for insightful and fruitful discussions. Especially, KT and SB would like to appreciate to Jose Beltr\'{a}n Jim\'{e}nez for giving insightful comments. KT is supported by Tokyo Tech Fund Hidetoshi Kusama Scholarship. SB is supported by “Agencia Nacional de Investigaci´on y Desarrollo” (ANID), Grant “Becas Chile postdoctorado al extranjero” No. 74220006.
\end{acknowledgments}

\appendix
\section{\label{sec:03:01}A reconstruction of Dirac-Bergmann analysis and degrees of freedom
}
Let us consider a first-order derivative point particle system:
\begin{equation}
L=L(\dot{q}^{i},q^{i})\,,
\label{PL}
\end{equation}
where $q^{i}:=q^{i}(t)$ are position variable, $\dot{q}^{i}:=\dot{q}^{i}(t)$ are velocity variable, and $t$ is time variable. The index $i$ run from $1$ to $n$. Then, the canonical momentum variables, denote $p_{i}(t)$s, are introduced through the first-order variation of the Lagrangian as follows:
\begin{equation}
\delta L = \left[\frac{\partial L}{\partial q^{i}}-\frac{d}{dt}\left(\frac{\partial L}{\partial \dot{q}^{i}}\right)\right]\delta q^{i}+\frac{d}{dt}\left[\left(\frac{\partial L}{\partial \dot{q}^{i}}\right)\delta q^{i}\right]
:= \left[\rm{EoM}\right]_{i}\delta q^{i}+\frac{d}{dt}\left[p_{i}\delta q^{i}\right]\,,
\label{FVofPL}
\end{equation}
where $\left[\rm{EoM}\right]_{i}$ denoted the equations of motion. 
The Lagrangian can be related to a Hamiltonian by performing the Legendre transformation as usual but in general, this correspondence does not determine uniquely. That is, the rank of the Hessian matrix
\begin{equation}
K_{ij}:=\frac{\partial p_{i}}{\partial \dot{q}^{i}}=\frac{\partial^{2}L}{\partial\dot{q}^{i}\partial\dot{q}^{j}}
\label{HesseMatrix}
\end{equation}
governs the uniqueness of the Legendre transformation. If the rank is equal to $n$ then the Hamiltonian is uniquely determined from the Lagrangian. However, if the rank is less than $n$ it is not the case: the system becomes a singular/degenerate system. In such a system, to reveal the time-development of the system, it needs a method to determine the unique Hamiltonian from the Lagrangian. This is nothing but the Dirac-Bergmann procedure~\cite{Dirac1950,Dirac1958,Bergmann1949,BergmannBrunings1949,Bergmann1950,AndersonBergmann1951}. 

Let us consider a case that the rank of the Hessian matrix is $n-r$ $(r\geq1)$. Then the number of $n-r$ velocity variables can be expressed by the same number of momentum variables by virtue of the implicit function theorem; there exists a set of functions $F^{a}$ such that $\dot{q}^{a}=F^{a}(q^{i},p_{a})$ $(a\in\{1,2,\cdots,n-r\})$, and the existence of the number of $r$ zero-eigenvalue vectors lead to the following relation~\cite{Sugano1989,Sugano1993}:
\begin{equation}
\left.\frac{\partial}{\partial\dot{q}^{\alpha}}\frac{\partial L}{\partial\dot{q}^{\beta}}\right|_{\dot{q}^{a}=F^{a}(q^{i},p_{a})}=0\,,
\label{DegenVV}
\end{equation}
where $\alpha$ and $\beta$ run form $n-r+1$ to $n$. This relation implies that the Lagrangian linearly includes the velocity variables $\dot{q}^{\alpha}$, therefore the Legendre transformation becomes as follows:
\begin{equation}
H=H_{0}+\dot{q}^{\alpha}\phi^{(1)}_{\alpha}\,,
\label{OriginalTotalHamiltonian}
\end{equation}
where $\phi^{(1)}_{\alpha}$ are set as $\phi^{(1)}_{\alpha}:=p_{\alpha}-f_{\alpha}(q^{i},p_{i})$ for some functions $f_{\alpha}(q^{i},p_{i})$. In the Lagrange formulation, $\phi^{(1)}_{\alpha}$ vanish as identity. However, in the Hamiltonian formulation, one needs to impose that those quantities vanish:
\begin{equation}
\phi^{(1)}_{\alpha}=p_{\alpha}-f_{\alpha}(q^{i},p_{i}):=0\,.
\label{OriginalPriConsts}
\end{equation}
These conditions are called primary constraint and restrict the whole phase space to a phase subspace denoted by $\mathfrak{C}^{(1)}:=\{p\in T^{*}\mathcal{P}|\phi^{(1)}_{\alpha}=0\}$, in which the time evolution has to proceed, where $\mathcal{P}$ is the configuration space of the system and $T^{*}\mathcal{P}$ is the whole phase space corresponding to the configuration space. To ensure this property, Eq~(\ref{OriginalPriConsts}) has to satisfy the so-called consistency conditions:
\begin{equation}
\dot{\phi}^{(1)}_{\alpha}=\{\phi^{(1)}_{\alpha},H_{T}\}\approx\{\phi^{(1)}_{\alpha},H_{0}\}+D^{(1)}_{\alpha\beta}\lambda^{\beta}:\approx0\,,
\label{OriginalPriConsistencyConds}
\end{equation}
where $H_{T}$ and $D^{(1)}_{\alpha\beta}$ are called total Hamiltonian and Dirac matrix, respectively, which are given as follows:
\begin{equation}
H_{T}:=H_{0}+\lambda^{\alpha}\phi^{(1)}_{\alpha}\,,\ \ \ D^{(1)}_{\alpha\beta}:=\{\phi^{(1)}_{\alpha},\phi^{(1)}_{\beta}\}\,,
\label{OriginalTotalHamiltonianAndDiracMatrix}
\end{equation}
and ``$\approx$' 
means that the equation is satisfied restricted to the phase sub-space $\mathfrak{C}^{(1)}$. In the above equation, the velocity variables $\dot{q}^{\alpha}$ are replaced by Lagrange multipliers $\lambda^{\alpha}$. This manipulation is possible since $\dot{q}^{\alpha}$ are undetermined due to the degeneracy of the Hessian matrix. 

Depending on the rank of the Dirac matrix $D^{(1)}$, a part of the multipliers are determined, but others remain arbitrary. Let us assume the rank as $r-r_{1}$. Then, performing the fundamental matrix transformations to $D^{(1)}$, there exists non-singular matrices $P^{(1)}$ and $Q^{(1)}$ such that $D'^{(1)}:=P^{(1)}D^{(1)}Q^{(1)}$. Note that it is free to choose the shape of $D'^{(1)}$ for ease of analysis. In this section, we take the shape of $D'^{(1)}$ as the standard form: $D'^{(1)}={\rm{diag}}(\tau^{(1)}_{1},\tau^{(1)}_{2},\cdots,\tau^{(1)}_{r-r_{1}},0,\cdots,0)$. Therefore, the multipliers of the same number to the rank are determined and the consistency conditions Eq~(\ref{OriginalPriConsistencyConds}) becomes as follows:
\begin{equation}
{P^{(1)}}_{\alpha}^{\beta}\{\phi^{(1)}_{\beta},H_{0}\}+D'^{(1)}_{\alpha\beta}\lambda'^{\beta}:\approx0\,,
\label{TransOriginalPriConsistencyConds}
\end{equation}
where ${\lambda'^{(1)}}^{\alpha}={{Q^{(1)}}^{-1}}^{\alpha}_{\beta}\lambda^{\beta}$. For the indices $a^{(1)}\in\{1,2,\cdots,r-r_{1}\}$, on one hand, the multipliers are determined as ${\lambda'}_{a^{(1)}}=-{\tau^{(1)}}_{a^{(1)}}{P^{(1)}}_{a^{(1)}}^{\beta}\{\phi^{(1)}_{\beta},H_{0}\}$. Notice that we do not sum over with respect to $a^{(1)}$. For the indices $\alpha^{(1)}\in\{r-r_{1}+1,r-r_{1}+2,\cdots,r\}$, on the other hand, new constraints, secondary constraints, appear:
\begin{equation}
\phi^{(2)}_{\alpha^{(1)}}:={P^{(1)}}_{\alpha^{(1)}}^{\beta}\{\phi^{(1)}_{\beta},H_{0}\}:\approx0\,.
\label{OriginalSecConsts}
\end{equation}
Then, the total Hamiltonian is arranged as follows:
\begin{equation}
H^{(1)}_{T}:=H^{(1)}_{0}+\lambda^{\alpha^{(1)}}\Phi^{(1)}_{\alpha^{(1)}}\,,
\label{OriginalPriHT}
\end{equation}
where $H^{(1)}_{0}$ and $\Phi^{(1)}_{\alpha^{(1)}}$ are defined as follows:
\begin{equation}
H^{(1)}_{0}:=H_{0}+\lambda^{a^{(1)}}\phi^{(1)}_{a^{(1)}}\,,\ \ \ \Phi^{(1)}_{\alpha^{(1)}}:=\phi^{(1)}_{\beta}{Q^{(1)}}^{\beta}_{\alpha^{(1)}}\,.
\label{OriginalPriH0AndPriConsts}
\end{equation}

Repeating the same procedure for the secondary constraints under the redefined total Hamiltonian Eq~(\ref{OriginalPriHT}), and if it gives rise to new constraints then repeat over the same process until all the multipliers are determined or the new constraint does not appear. Let us assume the process stops by $K$-steps. Then we obtain the final results; the constraints $\phi^{(1)}_{\alpha},\ \phi^{(2)}_{\alpha^{(1)}},\cdots,\ \phi^{(K)}_{\alpha^{(K-1)}}$ appear, the multipliers $\lambda^{a^{(1)}},\ \lambda^{a^{(2)}},\cdots,\ \lambda^{a^{(K)}}$ are determined, the redefined total Hamiltonian $H^{(K)}_{T}:=H^{(K)}_{0}+\lambda^{\alpha^{(K)}}\Phi^{(K)}_{\alpha^{(K)}}$ is derived, where $H^{(K)}_{0}:=H^{(K-1)}_{0}+\lambda^{\alpha^{(K)}}\phi^{(1)}_{\alpha^{(K)}}$ and $\Phi^{(K)}_{\alpha^{(K)}}:=\phi^{(K)}_{\beta}{Q^{(K)}}^{\beta}_{\alpha^{(K)}}$, and the dynamics of the system is restricted to the phase subspace $\mathfrak{C}^{(K)}:=\{p\in T^{*}\mathcal{P}|\phi^{(s)}_{\alpha_{s-1}}=0\}$ for all $s\in\{1,2,\cdots,K\}$, where $\alpha_{0}:=\alpha$. The multipliers $\lambda^{\alpha^{(K)}}$ remain arbitrary and the existence of such multipliers then implies that this system has the number of $\lambda^{\alpha^{(K)}}$ gauge Degrees of Freedom (gDoF)~\cite{Kimura1990,Sugano1990,Sugano1991,Kagraoka1991}. In such a case, the dynamics of the system does not uniquely determine without fixing the gDoF~\cite{Sugano1992}. 

The Dirac-Bergmann analysis reveals all constraints of the system. These constraints are classified into two classes: first-class and second-class. The former is defined as a set of constraints that are commutative with all other constraints in the phase subspace $\mathfrak{C}^{(K)}$ with respect to the Poisson bracket. Otherwise, the constraints are classified into second-class and the total number of second-class constraints is always even. Armed with this classification, an important theorem holds~\cite{Shanmugadhasan1973,Maskawa1976,Dominici1980,Dominici1982,Kyosuke2023}:\newline\newline
{\it For a symplectic form of the system: $\Omega=dq^{i}\wedge dp_{i}$ $(i\in\{1,2,\cdots,n\})$, it exists a canonical coordinate system such that $\Omega=dQ^{I}\wedge dP_{I}+d\Theta^{\alpha}\wedge d\Theta_{\alpha}+d\Xi^{a}\wedge d\Psi_{a}$ ($I\in\{1,2,\cdots,n-2u-v\}$; $\alpha\in\{1,2,\cdots,2u\}$; $a\in\{1,2,\cdots,v\}$), where $\Theta^{\alpha}$ and $\Theta_{\alpha}$s are composed only of all the $2u$ second-class constraints, $\Psi_{a}$ are composed only of all the $v$ first-class constraints.}\newline\newline
Since $\Theta^{\alpha}$, $\Theta_{\alpha}$, and $\Psi_{a}$ satisfy those consistency conditions, restricting $\Omega$ to $\mathfrak{C}^{(K)}$, we obtain $\Omega\approx dQ^{I}\wedge dP_{I}$. Therefore, the pDoF is the half number of the dimension of the phase subspace $\mathfrak{C}^{(K)}$, that is, ${\rm{pDoF}}=(2n-2u-2\times v)/2=n-u-v$.
This number is the main concept of this paper. The point is that to derive pDoF we just perform the Dirac-Bergmann analysis and count the total number of each class of constraints although it is generically difficult to find the explicit forms of $\Theta^{\alpha}$, $\Theta_{\alpha}$, and $\Psi_{a}$. 

The extension of these frameworks to field theories is achieved straightforwardly through the usual manipulations; just replacing the variables that describe the system by fields in terms of density variables although it needs a careful manipulation for spatial boundary terms as mentioned in Sec.~\ref{sec:03:02}. 

\section{\label{App01}PB-algebras of coincident $f(Q)$-gravity in $(n+1)$ - dimensional spacetime}
In the below calculations, all spatial boundary terms are neglected according to the prescription as discussed in Sec.~\ref{sec:03:02}. The assumed spatial boundary conditions are $N_{I}(t\,,\partial\Sigma_{t}):=0$ for each leaf $\Sigma_{t}$. 

The PB-algebras among the primary constraint densities $\phi^{(1)}_{A}$ $(A\in\{0,I,\varphi\};\  I\in\{1,2,\cdots,n\})$ and the density $\mathcal{H}_{0}$:
\begin{equation}
\begin{split}
\{\phi^{(1)}_{0}(x),\mathcal{H}_{0}(y)\}=
&
\left[-\mathcal{C}^{f(Q)}_{0}-\frac{\sqrt{h}}{N}\frac{N^{I}}{N}f''\left(\partial_{J}N^{J}\partial_{I}\varphi-\partial_{I}N^{J}\partial_{J}\varphi\right)\right]\delta^{(n)}(\vec{x}-\vec{y})\,,\\
\{\phi^{(1)}_{I}(x),\mathcal{H}_{0}(y)\}=
&
\left[-\mathcal{C}^{f(Q)}_{I}+\frac{1}{n-1}\frac{f''}{f'}\pi\partial_{I}\varphi\right]\delta^{(n)}(\vec{x}-\vec{y})\,,\\
\{\phi^{(1)}_{\varphi}(x),\mathcal{H}_{0}(y)\}=
&
\left[-\mathcal{H}'_{0}-\frac{1}{n-1}\frac{f''}{f'}\pi\partial_{I}N^{I}\right]\delta^{(n)}(\vec{x}-\vec{y})\,,
\end{split}
\label{}
\end{equation}
where $\mathcal{H}'_{0}$ is defined by
\begin{equation}
\begin{split}
\mathcal{H}'_{0}:=N\sqrt{h}
&
\left[-f''\left\{{^{(n)}Q}-\varphi+\frac{1}{h}\left(\frac{1}{f'}\right)^{2}\left(\pi^{IJ}\pi_{IJ}-\frac{1}{n-1}\pi^{2}\right)\right\}+\lc{D}_{I}\left\{f''\left({^{(n)}Q}^{I}-{^{(n)}\tilde{Q}}^{I}\right)\right\}\right.\\
&
\left.-\frac{1}{N}\frac{N^{I}}{N}f'''\left(\partial_{J}N^{J}\partial_{I}\varphi-\partial_{I}N^{J}\partial_{J}\varphi\right)\right]\,,
\end{split}
\label{}
\end{equation}
and $\mathcal{C}^{f(Q)}_{0}$ is computed as follows:
\begin{equation}
\begin{split}
\mathcal{C}^{f(Q)}_{0}:=
&
-\sqrt{h}\left[f'{{^{(n)}Q}}-\lc{D}_{I}\left\{f'\left({^{(n)}Q}^{I}-{{^{(n)}\tilde{Q}}}^{I}\right)\right\}+f-\varphi f'-\frac{1}{hf'}\left(\pi^{IJ}\pi_{IJ}-\frac{1}{n-1}\pi^{2}\right)\right]\,.
\end{split}
\label{}
\end{equation}
$\mathcal{C}^{f(Q)}_{I}$ does not change from Eq $(\ref{C and Ci})$ excepting the range of summations. The PB-algebras among two of the primary constraint densities $\phi^{(1)}_{A}$ $(A\in\{0,I,\varphi\};\ I\in\{1,2,\cdots,n\})$ and the density $\mathcal{H}_{0}$:
\begin{equation}
\begin{split}
\{\{\phi^{(1)}_{0},\mathcal{H}_{0}\}(x),\phi^{(1)}_{0}(y)\}=
&
\frac{\sqrt{h}}{N}\frac{N^{I}}{N}\frac{2}{N}f''\left(\partial_{J}N^{J}\partial_{I}\varphi-\partial_{I}N^{J}\partial_{J}\varphi\right)\delta^{(n)}(\vec{x}-\vec{y})\,,\\
\{\{\phi^{(1)}_{0},\mathcal{H}_{0}\}(x),\phi^{(1)}_{I}(y)\}=
&
\frac{1}{n-1}\frac{1}{N}\frac{f''}{f'}\pi\partial_{I}\varphi\delta^{(n)}(\vec{x}-\vec{y})\,,\\
\{\{\phi^{(1)}_{0},\mathcal{H}_{0}\}(x),\phi^{(1)}_{\varphi}(y)\}=
&
\left[\sqrt{h}f''\left\{{^{(n)}Q}-\varphi+\frac{1}{h}\left(\frac{1}{f'}\right)^{2}\left(\pi^{IJ}\pi_{IJ}-\frac{1}{n-1}\pi^{2}\right)\right\}-\frac{1}{n-1}\frac{1}{N}\frac{f''}{f'}\pi\partial_{I}N^{I}\right.\\
&
\left.-\frac{\sqrt{h}}{N}\frac{N^{I}}{N}f'''\left(\partial_{J}N^{J}\partial_{I}\varphi-\partial_{I}N^{J}\partial_{J}\varphi\right)\right]\delta^{(n)}(\vec{x}-\vec{y})\,.
\end{split}
\label{}
\end{equation}
\begin{equation}
\begin{split}
\{\{\phi^{(1)}_{I},\mathcal{H}_{0}\}(x),\phi^{(1)}_{0}(y)\}=
&
-\frac{1}{N}\frac{\sqrt{h}}{N}f''\left(\partial_{J}N^{J}\partial_{I}\varphi-\partial_{I}N^{J}\partial_{J}\varphi\right)\delta^{(n)}(\vec{x}-\vec{y})\,,\\
\{\{\phi^{(1)}_{I},\mathcal{H}_{0}\}(x),\phi^{(1)}_{J}(y)\}=
&
\frac{n}{2(n-1)}\frac{h}{N}\frac{(f'')^{2}}{f'}\partial_{I}\varphi\partial_{J}\varphi\delta^{(n)}(\vec{x}-\vec{y})\,,\\
\{\{\phi^{(1)}_{I},\mathcal{H}_{0}\}(x),\phi^{(1)}_{\varphi}(y)\}=
&
\left[\frac{\sqrt{h}}{N}f'''\left(\partial_{J}N^{J}\partial_{I}\varphi-\partial_{I}N^{J}\partial_{J}\varphi\right)+\frac{1}{n-1}\frac{1}{f'}\pi\partial_{I}\varphi\left(f'''-\frac{f''}{f'}\right)\right.\\
&
\left.-\frac{n}{2(n-1)}\frac{\sqrt{h}}{N}\frac{(f'')^{2}}{f'}\partial_{J}N^{J}\partial_{I}\varphi\right]\delta^{(n)}(\vec{x}-\vec{y})\,.
\end{split}
\label{}
\end{equation}
\begin{equation}
\begin{split}
\{\{\phi^{(1)}_{\varphi},\mathcal{H}_{0}\}(x),\phi^{(1)}_{0}(y)\}=
&
\left[\sqrt{h}f''\left\{{^{(n)}Q}-\varphi+\frac{1}{h}\left(\frac{1}{f'}\right)^{2}\left(\pi^{IJ}\pi_{IJ}-\frac{1}{n-1}\pi^{2}\right)-\frac{f'''}{f''}\partial_{I}\varphi\left({^{(n)}Q}^{I}-{^{(n)}\tilde{Q}}^{I}\right)\right\}\right.\\
&
\left.-\frac{2\sqrt{h}}{n-1}\frac{f''}{f'}\pi\partial_{I}N^{I}+\frac{1}{n-1}\frac{f''}{f'}\frac{1}{N}\pi\partial_{I}N^{I}-\frac{\sqrt{h}}{N}\frac{N^{I}}{N}f'''\left(\partial_{J}N^{J}\partial_{I}\varphi-\partial_{I}N^{J}\partial_{J}\varphi\right)\right]\delta^{(n)}(\vec{x}-\vec{y}),\\
\{\{\phi^{(1)}_{\varphi},\mathcal{H}_{0}\}(x),\phi^{(1)}_{I}(y)\}=
&
\left[-\frac{n}{2(n-1)}\frac{\sqrt{h}}{N}\frac{(f'')^{2}}{f'}\partial_{J}N^{J}\partial_{I}\varphi-\frac{1}{n-1}\frac{f'''}{(f')^{2}}\partial_{I}\varphi\right.\\
&
+\left.\frac{\sqrt{h}}{N}f'''\left(\partial_{J}N^{J}\partial_{I}\varphi-\partial_{I}N^{J}\partial_{J}\varphi\right)+\frac{1}{n-1}\frac{\pi}{f'}\left(f'''-\frac{f''}{f'}\right)\partial_{I}\varphi\right]\delta^{(n)}(\vec{x}-\vec{y}),\\
\{\{\phi^{(1)}_{\varphi},\mathcal{H}_{0}\}(x),\phi^{(1)}_{\varphi}(y)\}=
&
\left[-\mathcal{H}''_{0}+\frac{1}{n-1}\left(\frac{f''}{f'}\right)^{2}\pi\partial_{I}N^{I}-\frac{1}{n-1}\frac{1}{f'}\left(f''''-\frac{f''}{f'}\right)\partial_{I}N^{I}\right.\\
&
\left.+\frac{n}{2(n-1)}\sqrt{h}\frac{(f'')^{2}}{f'}\partial_{I}N^{I}\partial_{J}N^{J}\right]\delta^{(n)}(\vec{x}-\vec{y})\,,\\
\end{split}
\label{}
\end{equation}
where $\mathcal{H}''_{0}$ is defined as follows:
\begin{equation}
\begin{split}
\mathcal{H}''_{0}:=N\sqrt{h}
&
\left[-f'''\left({^{(n)}Q}-\varphi\right)+f''-\frac{1}{h}\frac{1}{N}\left(\frac{1}{f'}\right)^{2}\left(1-\frac{2}{f'}\right)\left(\pi^{IJ}\pi_{IJ}-\frac{1}{n-1}\pi^{2}\right)\right.\\
&
\left.+\lc{D}_{i}\left\{f'''\left({^{(n)}Q}^{I}-{^{(n)}\tilde{Q}}^{I}\right)\right\}-\frac{1}{N}\frac{N^{I}}{N}f''''\left(\partial_{J}N^{J}\partial_{I}\varphi-\partial_{I}N^{J}\partial_{J}\varphi\right)\right]\,.
\end{split}    
\label{}
\end{equation}
The PB-algebras among the primary constraint densities $\phi^{(1)}_{A}$ $(A\in\{0,I,\varphi\};\ I\in\{1,2,\cdots,n\})$, $A_{I}/B$, $C_{I}/B$, and the density $\mathcal{H}_{0}$:
\begin{equation}
\left\{\{\phi^{(1)}_{A},\mathcal{H}_{0}\}(x),\left(\frac{A_{I}}{B}\right)(y)\right\}=0\,\,,\ \ \ \left\{\{\phi^{(1)}_{A},\mathcal{H}_{0}\}(x),\left(\frac{C_{I}}{B}\right)(y)\right\}=0\,.
\label{}
\end{equation}
The PB-algebras among the primary constraint densities $\phi^{(1)}_{A}$ $(A\in\{0,I,\varphi\};\ I\in\{1,2,\cdots,n\})$, $A_{I}/B$, and $C_{I}/B$:
\begin{equation}
\left\{\left(\frac{A_{I}}{B}\right)(x),\phi^{(1)}_{0}(y)\right\}=0\,\,,\ \ \ \left\{\left(\frac{C_{I}}{B}\right)(x),\phi^{(1)}_{0}(y)\right\}=-\frac{1}{f''}\frac{1}{\partial_{I}N^{I}}f'''\partial_{I}\varphi\delta^{(n)}(\vec{x}-\vec{y})\,.
\label{}
\end{equation}
\begin{equation}
\left\{\left(\frac{A_{I}}{B}\right)(x),\phi^{(1)}_{J}(y)\right\}=0\,\,,\ \ \ \left\{\left(\frac{C_{I}}{B}\right)(x),\phi^{(1)}_{J}(y)\right\}=0\,.
\label{}
\end{equation}
\begin{equation}
\left\{\left(\frac{A_{I}}{B}\right)(x),\phi^{(1)}_{\varphi}(y)\right\}=0\,\,,\ \ \ 
\left\{\left(\frac{C_{I}}{B}\right)(x),\phi^{(1)}_{\varphi}(y)\right\}=N\frac{1}{\partial_{I}N^{I}}\frac{1}{f''}\left[\frac{(f''')^{2}}{f''}-f''''\right]\partial_{I}\varphi\delta^{(n)}(\vec{x}-\vec{y})\,.
\label{}
\end{equation}

\section{\label{App02}The explicit formulae of $\alpha$ and $\beta^{I}_{J}$}
\begin{equation}
\begin{split}
\alpha:=&\frac{1}{2}\frac{n}{n-1}\frac{h}{N}\frac{(f'')^{2}}{f'}+\frac{1}{n-1}\frac{1}{N}\frac{f'''}{f'}\frac{1}{\partial_{I}N^{I}}\pi\\
&+\frac{1}{\partial_{K}N^{K}}\left[\frac{\sqrt{h}}{N}\partial_{K}N^{K}+\frac{1}{n-1}\frac{1}{f'}\pi\left(f'''-\frac{f'''}{f'}\right)-\frac{n}{2(n-1)}\frac{\sqrt{h}}{N}\frac{(f'')^{2}}{f'}\partial_{K}N^{K}\right]\\
&+\frac{1}{\partial_{K}N^{K}}\left[-\frac{n}{2(n-1)}\frac{\sqrt{h}}{N}\frac{(f'')^{2}}{f'}\partial_{K}N^{K}-\frac{1}{n-1}\frac{f''}{(f')^{2}}+\frac{\sqrt{h}}{N}f'''\partial_{K}N^{K}-\frac{1}{n-1}\frac{1}{f'}\left(f''''-\frac{f''}{f'}\right)\partial_{K}N^{K}\right]\\
&-\frac{f'''}{\partial_{K}N^{K}}\left[\frac{\sqrt{h}}{N}\partial_{K}N^{K}\right]+\left(\frac{1}{\partial_{K}N^{K}}\right)^{2}N\frac{f''''}{f''}\{\phi^{(1)}_{0},\mathcal{H}_{0}\}\\
&+\left(\frac{1}{\partial_{K}N^{K}}\right)^{2}\left(\frac{f'''}{f''}\right)^{2}\{\{\phi^{(1)}_{0},\mathcal{H}_{0}\},\phi^{(1)}_{0}\}+\left(\frac{1}{\partial_{K}N^{K}}\right)^{2}\frac{f'''}{f''}N\{\{\phi^{(1)}_{0},\mathcal{H}_{0}\},\phi^{(1)}_{\varphi}\}+\left(\frac{1}{\partial_{K}N^{K}}\right)^{2}\frac{f'''}{f''}N\{\{\phi^{(1)}_{\varphi},\mathcal{H}_{0}\},\phi^{(1)}_{0}\}\\
&+\left(\frac{1}{\partial_{K}N^{K}}\right)^{2}\{\{\phi^{(1)}_{\varphi},\mathcal{H}_{0}\},\phi^{(1)}_{\varphi}\}
\end{split}
\label{}
\end{equation}
\begin{equation}
\beta^{I}_{J}:=-\frac{1}{\partial_{K}N^{K}}\frac{\sqrt{h}}{N}f'''\partial_{J}N^{I}\,.
\label{}
\end{equation}
\bibliography{HAofCFQ}

\providecommand{\noopsort}[1]{}\providecommand{\singleletter}[1]{#1}%
\begin{thebibliography}{10}

\bibitem{Einstein1928}
A.~Einstein.
\newblock Riemann-geometrie mit aufrechterhaltung des begriffes des
  fernparallelismus.
\newblock {\em Preussische Akademie der Wissenschaften, Phys.Math. Klasse,
  Sitzungsberichte.}, page 217, 1928.

\bibitem{Bahamonde:2021gfp}
S.~Bahamonde, K.~F. Dialektopoulos, C.~Escamilla-Rivera, G.~Farrugia, V.~Gakis,
  M.~Hendry, M.~Hohmann, J.~S. Levi, J.~Mifsud, and E.~D. Valentino.
\newblock {Teleparallel gravity: from theory to cosmology}.
\newblock {\em Rept. Prog. Phys.}, 86(2):026901, 2023.

\bibitem{Nester:1998mp}
J.~M. Nester and Hwei-Jang Yo.
\newblock {Symmetric teleparallel general relativity}.
\newblock {\em Chin. J. Phys.}, 37:113, 1999.

\bibitem{Jimenez2019}
J.~B. Jimenez, L.~Heisenberg, and T.~S. Koivisto.
\newblock The geometrical trinity of gravity.
\newblock {\em Universe}, 5 (2019) no.7, 173, 2019.

\bibitem{Heisenberg:2018vsk}
Lavinia Heisenberg.
\newblock {A systematic approach to generalisations of General Relativity and
  their cosmological implications}.
\newblock {\em Phys. Rept.}, 796:1--113, 2019.

\bibitem{Buchdahl1970}
H.~A. Buchdahl.
\newblock Non-linear lagrangians and cosmological theory.
\newblock {\em MNRAS}, 150,1:1, 1970.

\bibitem{Jimenez2020}
J.~B. Jimenez, L.~Heisenberg, T.~Koivisto, and S.~Pekar.
\newblock Cosmology in $f({Q})$ geometry.
\newblock {\em Phys.Rev.D}, 101:103507, 2020.

\bibitem{Planck18}
Planck Collaboration:~N. Aghanim and {\it et\ al.}
\newblock Planck 2018 results. vi. cosmological parameters.
\newblock {\em A\&A 641, A}, 6, 2020.

\bibitem{Bessa2022}
P.~Bessa, M.~Campista, and A.~Bernui.
\newblock Observational constraints on starobinsky $f({R})$ cosmology from
  cosmic expansion and structure growth data.
\newblock {\em EPJC}, 82, 2022.

\bibitem{Stelle1978}
K.~S. Stelle.
\newblock Classical gravity with higher derivatives.
\newblock {\em Gen Relat Gravit}, 9:353, 1978.

\bibitem{Starobinsky1980}
A.~A. Starobinsky.
\newblock A new type of isotropic cosmological models without singularity.
\newblock {\em Phys.Lett.B 91}, 91:99, 1980.

\bibitem{Dirac1950}
P.~M.~A. Dirac.
\newblock Lectures on quantum mechanics.
\newblock {\em Can.J.Math.}, 2:129, 1950.

\bibitem{Dirac1958}
P.~M.~A. Dirac.
\newblock Generalized {H}amiltonian dynamics.
\newblock {\em Proc.R.Soc.London Ser.}, A 246:326, 1958.

\bibitem{Bergmann1949}
P.~G. Bergmann.
\newblock Non-linear field theories.
\newblock {\em Phys.Rev.}, 75:680, 1949.

\bibitem{BergmannBrunings1949}
P.~G. Bergmann and J.~H.~M. Brunings.
\newblock Non-linear field theories {II}. {C}anonical equations and
  quantization.
\newblock {\em Rev.Mod.Phys.}, 21:480, 1949.

\bibitem{Bergmann1950}
P.~G. Bergmann, R.~Penfield, R.~Schiller, and H.~Zatzkis.
\newblock The {H}amiltonian of the general theory of relativity with
  electromagnetic field.
\newblock {\em Phys.Rev.}, 80:81, 1950.

\bibitem{AndersonBergmann1951}
J.~L. Anderson and P.~G. Bergmann.
\newblock Constraints in covariant field theories.
\newblock {\em Phys.Rev.}, 83:1018, 1951.

\bibitem{Weitzenboh1923}
R.~Weitzenboch.
\newblock Invarianten theorie.
\newblock {\em Nordhoff, Groningen}, page 320, 1923.

\bibitem{Blagojevic2000}
M.~Blagojevic and I.~A. Nikolic.
\newblock Hamiltonian structure of the teleparallel formulation of {GR}.
\newblock {\em Phys.Rev.D}, 62:024021, 2000.

\bibitem{Maluf2001}
J.~W. Maluf and J.~F. da~Rocha-Neto.
\newblock Hamiltonian formulation of general relativity in the teleparallel
  geometry.
\newblock {\em Phys.Rev.D}, 64:084014, 2001.

\bibitem{Ferraro2016}
R.~Ferraro and M.~J. Guzmán.
\newblock Hamiltonian formulation of teleparallel gravity.
\newblock {\em Phys.Rev.D}, 94:104045, 2016.

\bibitem{Blagojevic2000-2}
M.~Blagojevic and M.~Vasilic.
\newblock Gauge symmetries of the teleparallel theory of gravity.
\newblock {\em Class.Quant.Grav.}, 17:3785, 2000.

\bibitem{Li2011}
M.~Li, Rong-Xin Miao, and Yan-Gang Miao.
\newblock Degrees of freedom of $f({T})$ gravity.
\newblock {\em JHEP}, 1107:108, 2011.

\bibitem{Ong2013}
Y.~C. Ong, K.~Izumi, J.~M. Nester, and P.~Chen.
\newblock Problems with propagation and time evolution in $f({T})$ gravity.
\newblock {\em Phys.Rev.D}, 88:024019, 2013.

\bibitem{Ferraro2018}
R.~Ferraro and M.~J. Guzmán.
\newblock Hamiltonian formalism for $f({T})$ gravity.
\newblock {\em Phys.Rev.D}, 97:104028, 2018.

\bibitem{Blagojevic2020}
M.~Blagojević and J.~M. Nester.
\newblock Local symmetries and physical degrees of freedom in $f({T})$ gravity:
  a dirac hamiltonian constraint analysis.
\newblock {\em Phys.Rev.D}, 102:064025, 2020.

\bibitem{Liang2017}
D.~Liang, Y.~Gong, S.~Hou, and Y.~Liu.
\newblock Polarizations of gravitational waves in $f({R})$ gravity.
\newblock {\em Phys.Rev.D}, 95:104034, 2017.

\bibitem{Blixt2021}
D.~Blixt, María-José Guzmán, M.~Hohmann, and C.~Pfeifer.
\newblock Review of the hamiltonian analysis in teleparallel gravity.
\newblock {\em Int.J.Geom.Methods\ Mod.Phys.}, Vol. 18, No. supp01:2130005,
  2021.

\bibitem{Golovnev:2018wbh}
A.~Golovnev and T.~Koivisto.
\newblock {Cosmological perturbations in modified teleparallel gravity models}.
\newblock {\em JCAP}, 11:012, 2018.

\bibitem{Bahamonde:2022ohm}
S.~Bahamonde, K.~F. Dialektopoulos, M.~Hohmann, J.~S. Levi, C.~Pfeifer, and
  E.~N. Saridakis.
\newblock {Perturbations in non-flat cosmology for f(T) gravity}.
\newblock {\em Eur. Phys. J. C}, 83(3):193, 2023.

\bibitem{Jimenez2018}
J.~B. Jimenez, L.~Heisenberg, and T.~Koivisto.
\newblock Coincident general relativity.
\newblock {\em Phys.Rev.D}, 98:044048, 2018.

\bibitem{Jimenez2022}
J.~B. Jimenez and T.~S. Koivisto.
\newblock Lost in translation: {T}he abelian affine connection (in the
  coincident gauge).
\newblock {\em Int. J. Geom. Methods Mod. Phys}, Vol. 19, No. 07:2250108, 2022.

\bibitem{Blixt2023}
D.~Blixt, A.~Golovnev, Maria-Jose Guzman, and R.~Maksyutov.
\newblock {Geometry and covariance of symmetric teleparallel theories of
  gravity}.
\newblock 2023.
\newblock arXiv:2306.09289 [gr-qc].

\bibitem{Katsuragawa2022}
K.~Hu, T.~Katsuragawa, and T.~Qiu.
\newblock {ADM} formulation and hamiltonian analysis of $f({Q})$ gravity.
\newblock {\em Phys.Rev.D}, 106:044025, 2022.

\bibitem{Hu:2023gui}
K.~Hu, M.~Yamakoshi, T.~Katsuragawa, S.~Nojiri, and Q.~Taotao.
\newblock {Nonpropagating ghost in covariant f(Q) gravity}.
\newblock {\em Phys. Rev. D}, 108(12):124030, 2023.

\bibitem{Fabio2023}
F.~D'Ambrosio, L.~Heisenberg, and S.~Zentarra.
\newblock {Hamiltonian Analysis of $f({Q})$ Gravity and the Failure of the
  Dirac-Bergmann Algorithm for Teleparallel Theories of Gravity}.
\newblock {\em arXiv:2308.02250 [gr-qc]}, 2023.

\bibitem{York1972}
J.~W. York.
\newblock Role of conformal three-geometry in the dynamics if gravitation.
\newblock {\em Phys.Rev.Lett.}, 28:1082, 1972.

\bibitem{GibbonsHawking1977}
G.~W. Gibbons and S.~W. Hawking.
\newblock Action integrals and partition functions in quantum gravity.
\newblock {\em Phys.Rev.D}, 15:2752, 1977.

\bibitem{York1986}
J.~W. York.
\newblock Boundary terms in the action principle of general relativity.
\newblock {\em Found.Phys.}, 16:249, 1986.

\bibitem{HawkingHorowitz1996}
S.~W. Hawking and Gary~T. Horowitz.
\newblock The gravitational hamiltonian, action, entropy, and surface terms.
\newblock {\em Class.Quant.Grav.}, 13:1487, 1996.

\bibitem{Sundermeyer:1982}
K.~Sundermeyer.
\newblock {\em Constrained Dynamics. Lecture Notes in Physics.}
\newblock Springer, 1982.

\bibitem{ADM1959}
R.~Arnowitt, S.~Deser, and C.~W. Misner.
\newblock Dynamical structure and definition of energy in general relativity.
\newblock {\em Phys.Rev.}, 116:1322, 1959.

\bibitem{ADM1960}
R.~Arnowitt, S.~Deser, and C.~W. Misner.
\newblock Canonical variables for general relativity.
\newblock {\em Phys.Rev.}, 117:1595, 1960.

\bibitem{Baez1994}
J.~Baez~(UC Riverside) and J.~P. Muniain~(UC Riverside).
\newblock {\em Gauge Fields, Knots and Gravity}.
\newblock World Scientific, 1994.

\bibitem{Nakahara2003}
M.~Nakahara.
\newblock {\em Geometry, Topology and Physics, Second Edition (Graduate Student
  Series in Physics)}.
\newblock CRC Press, 2003.

\bibitem{Carroll1997}
S.~M. Carroll.
\newblock Lecture notes on general relativity.
\newblock {\em NSF-ITP/97-147}, 1997.

\bibitem{Hehl1995}
F.~W. Hehl, J.~D. McCrea, E.~W. Mielke, and Y.~Ne'eman.
\newblock Metric-affine gauge theory of gravity: {F}ield equations, {N}oether
  identities, world spinors, and breaking of dilation invariance.
\newblock {\em Phys.Rept.}, vol. 258:1, 1995.

\bibitem{Hilbert1915}
D.~Hilbert.
\newblock Die grundlagen der physik . (erste mitteilung.).
\newblock {\em Nachrichten von der Gesellschaft der Wissenschaften zu
  Göttingen – Mathematisch-Physikalische Klasse (in German)}, 3:395, 1915.

\bibitem{Einstein1916}
A.~Einstein.
\newblock Hamilton's principle and the general theory of relativity.
\newblock {\em Sitzungsber.Preuss.Akad.Wiss.Berlin (Math.Phys)}, page 1111,
  1916.

\bibitem{Keisuke2023}
K.~Izumi, K.~Shimada, K.~Tomonari, and M.~Yamaguchi.
\newblock {Boundary conditions for constraint systems in variational
  principle}.
\newblock {\em PTEP}, 2023(10):103E03, 2023.

\bibitem{Kyosuke2023}
K.~Tomonari.
\newblock {On the well-posed variational principle in degenerate point particle
  systems using embeddings of the symplectic manifold}.
\newblock {\em PTEP}, 2023(6):063A05, 2023.

\bibitem{Padmanabhan2006}
A.~Mukhopadhyay and T.~Padmanabhan.
\newblock Holography of gravitational action functionals.
\newblock {\em Phys.Rev.D}, 74:124023, 2006.

\bibitem{Ostrogradsky1850}
M.~V. Ostrogradski.
\newblock Mémoires sur les équations différentielles, relatives au problème
  des isopérimètres.
\newblock {\em Mem.Acad.St.Petersbourg}, IV 4:385, 1850.

\bibitem{Woodard2015}
R.~P. Woodard.
\newblock The theorem of ostrogradski.
\newblock {\em Scholarpedia}, 10:32243, 2015.

\bibitem{Langlois2016}
D.~Langlois and K.~Noui.
\newblock Degenerate higher derivative theories beyond horndeski: evading the
  ostrogradski instability.
\newblock {\em JCAP}, 02(2016)034, 2016.

\bibitem{Crisostomi2016}
M.~Crisostomi, K.~Koyama, and G.~Tasinato.
\newblock Extended scalar-tensor theories of gravity.
\newblock {\em JCAP}, 04(2016)044, 2016.

\bibitem{Achour2016}
J.~B. Achour, D.~Langlois, and K.~Noui.
\newblock Degenerate higher order scalar-tensor theories beyond horndeski and
  disformal transformations.
\newblock {\em Phys.Rev.D 93}, 93:124005, 2016.

\bibitem{Sugano1989}
Y.~Saito, R.~Sugano, T.~Ohta, and T.~Kimura.
\newblock A dynamical structure of singular {L}agrangian system with higher
  derivatives.
\newblock {\em J.Math.Phys.}, 30:1122, 1989.

\bibitem{Sugano1993}
Y.~Saito, R.~Sugano, T.~Ohta, and T.~Kimura.
\newblock Addendum to a dynamical structure of singular {L}agrangian system
  with higher derivatives.
\newblock {\em J.Math.Phys.}, 34:3775, 1993.

\bibitem{Pons1989}
J.~M. Pons.
\newblock Ostrogradski's theorem for higher-order singular {L}agrangians.
\newblock {\em Lett.Math.Phys.}, 17:181, 1989.

\bibitem{Jabbari1999}
M.M. Sheikh-Jabbari and A.~Shirzada.
\newblock {Boundary Conditions as Dirac Constraints}.
\newblock {\em Eur. Phys. J.}, C19:383, 2001.

\bibitem{Alhamawi2019}
A.~Alhamawi and R.~Alhamawi.
\newblock Generalized {G}ibbons-{H}awking-{Y}ork term for $f({R})$ gravity.
\newblock {\em J.Phys.:Conf.Ser.}, 1294:032032, 2019.

\bibitem{Kimura1990}
R.~Sugano and T.~Kimura.
\newblock Gauge transformations for dynamical systems with first- and
  second-class constraints.
\newblock {\em Phys.Rev.D}, 41:1247, 1990.

\bibitem{Sugano1990}
R.~Sugano and T.~Kimura.
\newblock Classification of gauge groups in terms of algebraic structure of
  first class constraints gauge transformations.
\newblock {\em J.Math.Phys.}, 31:2337, 1990.

\bibitem{John1972}
J.~O'Hanlon.
\newblock {Intermediate-Range Gravity: A Generally Covariant Model}.
\newblock {\em Phys.Rev.Lett.}, 29:137, 1972.

\bibitem{Teyssandier1983}
P.~Teyssandier and Ph. Tourrenc.
\newblock {The Cauchy problem for the R+R2 theories of gravity without
  torsion}.
\newblock {\em J.Math.Phys.}, 24:2793, 1983.

\bibitem{Kasper2007}
K.~Peeters.
\newblock {\em SPIN-06/46, ITP-UU-06/56}, 2007.

\bibitem{Fabio2020}
F.~D'Ambrosio, M.~Garg, L.~Heisenberg, and S.~Zentarra.
\newblock {ADM} formulation and {H}amiltonian analysis of coincident general
  relativity.
\newblock arXiv:2007.03261 [gr-qc], 2020.

\bibitem{Castellani:1981us}
L.~Castellani.
\newblock {Symmetries in Constrained Hamiltonian Systems}.
\newblock {\em Annals Phys.}, 143:357, 1982.

\bibitem{Kaparulin:2012px}
D.~S. Kaparulin, S.~L. Lyakhovich, and A.~A. Sharapov.
\newblock {Consistent interactions and involution}.
\newblock {\em JHEP}, 01:097, 2013.

\bibitem{ErrastiDiez:2020dux}
Ver\'onica Errasti~D\'\i{}ez, Markus Maier, Julio~A. M\'endez-Zavaleta, and
  Mojtaba Taslimi~Tehrani.
\newblock {Lagrangian constraint analysis of first-order classical field
  theories with an application to gravity}.
\newblock {\em Phys. Rev. D}, 102:065015, 2020.

\bibitem{ErrastiDiez:2023gme}
Ver\'onica Errasti~D\'\i{}ez, Markus Maier, and Julio~A. M\'endez-Zavaleta.
\newblock {Constraint characterization and degree of freedom counting in
  Lagrangian field theory}.
\newblock 10 2023.

\bibitem{Gomes:2023hyk}
D\'ebora~Aguiar Gomes, Jose~Beltr\'an Jim\'enez, and Tomi~S. Koivisto.
\newblock {General Parallel Cosmology}.
\newblock 9 2023.

\bibitem{Gomes:2023tur}
D\'ebora~Aguiar Gomes, Jose~Beltr\'an Jim\'enez, Alejandro~Jim\'enez Cano, and
  Tomi~S. Koivisto.
\newblock {On the pathological character of modifications of Coincident General
  Relativity: Cosmological strong coupling and ghosts in $f(\mathbb{Q})$
  theories}.
\newblock 11 2023.

\bibitem{Heisenberg:2023wgk}
Lavinia Heisenberg, Manuel Hohmann, and Simon Kuhn.
\newblock {Cosmological teleparallel perturbations}.
\newblock 11 2023.

\bibitem{Biswas2006}
T.~Biswas, A.~Mazumdar, and W.~Siegel.
\newblock Bouncing universes in string-inspired gravity.
\newblock {\em JCAP}, 0603(2006)009, 2006.

\bibitem{Capozziello2022}
S.~Capozziello and F.~Bajardi.
\newblock Non-local gravity cosmology: an overview.
\newblock {\em Int.J.Mod.Phys.D}, Vol.31,No.06,2230009, 2022.

\bibitem{Koshelev2023}
A.~S. Koshelev, K.~S. Kumar, and Alexei~A. Starobinsky.
\newblock Cosmology in nonlocal gravity.
\newblock {\em arXiv:2305.18716 [hep-th]}, 2023.

\bibitem{Joshi2022}
P.~Joshi and S.~Panda.
\newblock Hamiltonian analysis of nonlocal f({R}) gravity models.
\newblock {\em EPJC}, 82:601, 2022.

\bibitem{Souza2004}
J.~Moses P.~N.~de Souza, R.~Fateman and C.~Yapp.
\newblock {\em The Maxima book}.
\newblock https://maxima.sourceforge.io, 2004.

\bibitem{Sugano1991}
R.~Sugano and Y.~Kagraoka.
\newblock Extension to velocity dependent gauge transformations {I}. general
  form of velocity the generator.
\newblock {\em Z.Phys.C Particles and Fields}, 52:437, 1991.

\bibitem{Kagraoka1991}
R.~Sugano and Y.~Kagraoka.
\newblock Extension to velocity dependent gauge transformations {II}.
  properties of velocity dependent.
\newblock {\em Z.Phys.C Particles and Fields}, 52:443, 1991.

\bibitem{Sugano1992}
R.~Sugano, Y.~Kagraoka, and T.~Kimura.
\newblock Gauge transformations and gauge-fixing conditions in constraint
  systems.
\newblock {\em J.Math.Phys. A}, 7:62, 1992.

\bibitem{Shanmugadhasan1973}
S.~Shanmugadhasan.
\newblock Canonical formalism for degenerate lagrangians.
\newblock {\em J.Math.Phys.}, 14:67, 1973.

\bibitem{Maskawa1976}
T.~Maskawa and H.~Nakajima.
\newblock Singular {L}agrangian and the {D}irac-{F}adeev method: {E}xistence
  theorem of constraints in 'standard form'.
\newblock {\em Prog.Theor.Phys.}, 56:1295, 1976.

\bibitem{Dominici1980}
D.~Dominici and J.~Gomis.
\newblock Poincare-{C}artan integral invariant and canonical transformations
  for singular lagrangians.
\newblock {\em J.Math.Phys.}, 21:2124, 1980.

\bibitem{Dominici1982}
D.~Dominici.
\newblock Poincare-{C}artan integral invariant and canonical transformations
  for singular lagrangians: An addendum.
\newblock {\em J.Math.Phys.}, 23:256, 1982.

\end{thebibliography}
\bibliographystyle{unsrt}
\end{document}